\documentclass[%
 reprint,
superscriptaddress,
 amsmath,amssymb,
 aps,
prx,
]{revtex4-2}

\usepackage[utf8]{inputenc}
\usepackage[T1]{fontenc}

\usepackage{graphicx}
\usepackage[caption=false]{subfig}  
\usepackage{color}
\usepackage{natbib}
\usepackage{amsmath}
\usepackage{mathtools}
\usepackage{bbm}
\usepackage{physics}
\usepackage[normalem]{ulem}
\usepackage[ruled,vlined]{algorithm2e}

\usepackage[caption=false]{subfig}
\usepackage[hypertexnames=false]{hyperref}

\usepackage{enumitem}  
\usepackage{cleveref}

\usepackage{booktabs}
\usepackage{array}
\newcolumntype{L}[1]{>{\raggedright\let\newline\\\arraybackslash\hspace{0pt}}m{#1}}
\newcolumntype{C}[1]{>{\centering\let\newline\\\arraybackslash\hspace{0pt}}m{#1}}
\newcolumntype{R}[1]{>{\raggedleft\let\newline\\\arraybackslash\hspace{0pt}}m{#1}}

\crefname{figure}{Fig.}{Figures}
\crefname{equation}{Eq.}{Equations}
\crefname{section}{Section}{Sections}
\crefname{appendix}{Appendix}{Appendices}

\newcommand{\trise}{t_\textnormal{r}}
\newcommand{\im}{\textnormal{i}}
\newcommand{\naturalE}{\textnormal{e}}

\newcommand{\expect}[1]{\langle #1 \rangle}

\newcommand{\tfinal}{t_\textnormal{f}}

\renewcommand{\subsection}[1]{\paragraph{\textbf{#1}}}

\begin{document}

\begin{abstract}

While quantum circuits are reaching impressive widths in the hundreds of qubits, their depths have not been able to keep pace. In particular, cloud computing gates on multi-qubit, fixed-frequency superconducting chips continue to hover around the 1\% error range, contrasting with the progress seen on carefully designed two-qubit chips, where error rates have been pushed towards 0.1\%. Despite the strong impetus and a plethora of research, experimental demonstration of error suppression on these multi-qubit devices remains challenging, primarily due to the wide distribution of qubit parameters and the demanding calibration process required for advanced control methods. Here, we achieve this goal, using a simple control method based on multi-derivative, multi-constraint pulse shaping,  which acts simultaneously against multiple error sources. Our approach establishes a two to fourfold improvement on the default calibration scheme, demonstrated on four qubits on the IBM Quantum Platform with limited and intermittent access, enabling these large-scale fixed-frequency systems to fully take advantage of their superior coherence times. The achieved CNOT fidelities of 99.7(1)\% on those publically available qubits come from both coherent control error suppression and accelerated gate time.
\end{abstract}

\title{Experimental error suppression in Cross-Resonance gates \\ via multi-derivative pulse shaping}

\author{Boxi Li}
\email{b.li@fz-juelich.de}
\affiliation{Forschungszentrum Jülich, Institute of Quantum Control (PGI-8), D-52425 Jülich, Germany}
\affiliation{Institute for Theoretical Physics, University of Cologne, D-50937 Cologne, Germany}
\author{Tommaso Calarco}
\affiliation{Forschungszentrum Jülich, Institute of Quantum Control (PGI-8), D-52425 Jülich, Germany}
\affiliation{Institute for Theoretical Physics, University of Cologne, D-50937 Cologne, Germany}
\affiliation{Dipartimento di Fisica e Astronomia, Università di Bologna, 40127 Bologna, Italy}
\author{Felix Motzoi}
\email{f.motzoi@fz-juelich.de}
\affiliation{Forschungszentrum Jülich, Institute of Quantum Control (PGI-8), D-52425 Jülich, Germany}

\maketitle

\section{Introduction}
\begin{figure*}[t]
    \centering
    \includegraphics[width=0.95\textwidth]{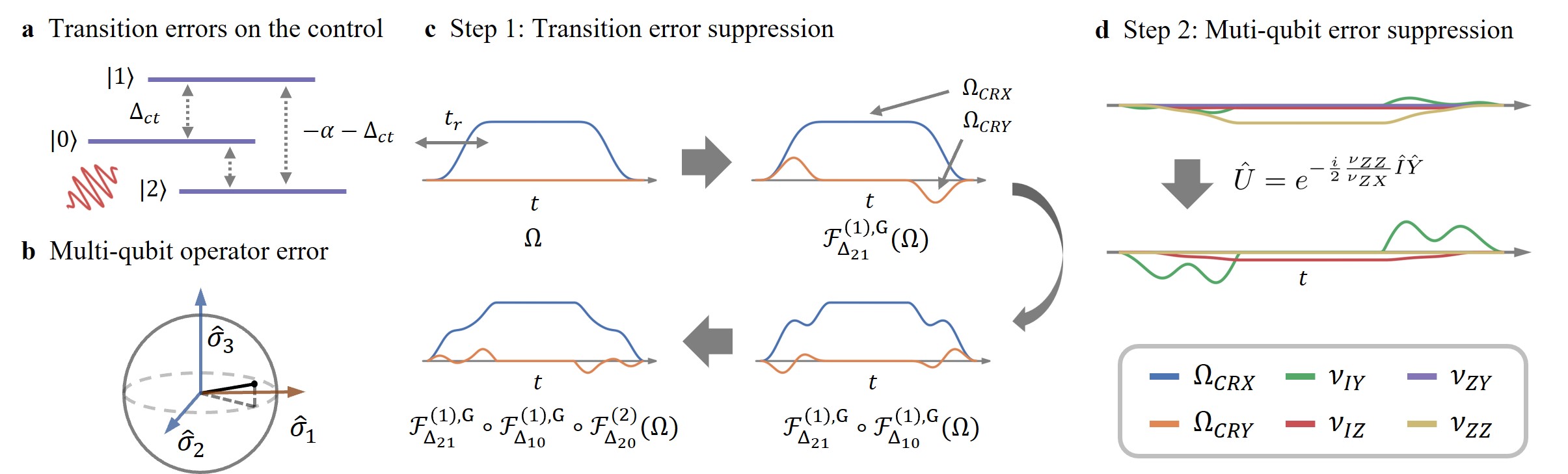}
    \caption{
    Illustration of the dominant coherent errors in the Cross-Resonance gate and the proposed pulse schemes to correct them.
    {\bf a.}~
    Transition errors in the rotating frame for a Transmon qubit driven off-resonantly, with $\Delta_{ct}$ the detuning and $\alpha$ the anharmonicity.
    {\bf b.}~
    Two-qubit errors of the CR gate. The three axes \{$\hat{\sigma}_1,\hat{\sigma}_2,\hat{\sigma}_3$\} represent either \{$\hat{\sigma}_{ZX},\hat{\sigma}_{ZY},\hat{\sigma}_{ZZ}$\} or \{$\hat{\sigma}_{IX},\hat{\sigma}_{IY},\hat{\sigma}_{IZ}$\}.
    The Hamiltonians $ZX$ and $IX$ (brown) commute and are defined as the ideal dynamics, while the others are considered errors (blue).
    {\bf c.}~
    Schematic illustration of the recursive \textsc{drag} pulse that suppresses different error transitions on the control Transmon.
    {\bf d.}~
    Schematic illustration of different multi-qubit errors in the effective frame during the CR operation, and the transformations of the error terms.
    The remaining $IY$ and $IZ$ errors are compensated for by an $IY$-\textsc{drag} pulse on the target qubit and the detuning of the CR drive.
    }
    \label{fig:overview}
\end{figure*}

Superconducting qubits have experienced significant improvement in the last decade, reaching the error correction threshold~\cite{Zhao2022Realization,Acharya2023Suppressing} and been used to study nontrivial quantum phenomena~\cite{Satzinger2021Realizing,Andersen2023NonAbelian}.
Additionally, quantum devices have become more accessible outside research labs, with cloud-based platforms like the IBM Quantum platform~\cite{IBM} providing access to multi-qubit devices, on which near-term quantum computing applications with error mitigation have been demonstrated~\cite{Kim2023Evidence}.
Although very high gate fidelity has been achieved on isolated chips~\cite{Kandala2021Demonstration,Wei2022Hamiltonian}, gate performance on scalable, publicly available multi-qubit devices is still bottlenecked, especially for two-qubit operations~\cite{Willsch2017Gateerror}.
These control imperfections not only limit the fidelity and depth of quantum circuits, but also give rise to correlated errors that propagate across the qubit lattice, sabotaging quantum error correction~\cite{Ghosh2013Understanding}, making error mitigation and benchmarking more challenging \cite{McEwen2021Removing, Miao2023Overcoming,Varbanov2020Leakage,Bultink2020Protecting}.

The Cross-Resonance (CR) gate is one of the most widely used two-qubit entangling gates for superconducting qubits, using microwave controls and avoiding noisy flux lines~\cite{Paraoanu2006Microwaveinduced,DeGroot2010Selective,Rigetti2010Fully,Groszkowski2011Tunable,Sheldon2016Procedure}.
It is the default gate on most devices provided by IBM and has found applications in high-quality circuit implementation, parity measurement, and state preparation~\cite{Takita2016Demonstration,Takita2017Experimental,Kandala2021Demonstration}.
While the absence of flux control lines extends qubit coherence time, it limits qubit tunability and necessitates weak coupling between qubits.
Consequently, achieving fast two-qubit gates requires a strong drive, which often leads to coherent errors due to non-adiabatic dynamics.
In practice, limiting the drive amplitude and a long pulse ramping time are used to prevent undesired dynamics,
including off-resonant transitions introduced by the drive~\cite{Malekakhlagh2022Mitigating,Tripathi2019Operation,Wei2024Characterizing}
and unwanted dynamics in the effective qubits' subspace ~\cite{Magesan2020Effective,Malekakhlagh2020Firstprinciples,Tripathi2019Operation}
(see \cref{fig:overview}a and \cref{fig:overview}b).

To circumvent control errors while maintaining substantial coupling strength, a combination of careful qubit parameter engineering and advanced control schemes has been employed. 
With these techniques, the best CR gate infidelity reported lies between 0.1\% and 0.3\%~\cite{Kandala2021Demonstration, Wei2022Hamiltonian}.
However, extending these advancements to scalable multi-qubit devices proves challenging.
For instance, on the 127-qubit chip {\it ibm\_brisbane}, the best echoed CR gate has an error of 0.35\%, while the median is only around 0.8\%, considerably higher than their coherence limit (median T1$\approx$$200\mu$s and T2$\approx$$135\mu$s)~\cite{IBM}.
An important factor contributing to the challenge lies in the intentional distribution of qubit parameters over a wide range, a design choice aimed at mitigating cross-talk~\cite{Hertzberg2021Laserannealing,Zhang2022Highperformance,Zhao2022Quantum}.
This uncertainty in the qubit parameters also stems unavoidably from the inhomogeneity in the fabrication process~\cite{Tolpygo2015Fabrication, Hertzberg2021Laserannealing}.
Therefore, designing an efficient control scheme that operates seamlessly across diverse parameter regimes is essential for achieving optimal performance across hundreds of qubits.

In this work, we devise and test a simple and scalable control scheme for CR gates that counteracts all the aforementioned control errors, following the ideal of the well-known Derivative Removal by Adiabatic Gate (\textsc{drag}) method~\cite{Motzoi2009Simple,Gambetta2011Analytic,Motzoi2013Improving,Theis2018Counteracting}.
For the transition error, we demonstrate that previously suggested single-derivative \textsc{drag} correction~\cite{Malekakhlagh2022Mitigating} is insufficient for typical parameters in multi-qubit devices, where multiple transition errors are present.
We introduce a novel recursive multi-derivative \textsc{drag} pulse, considering all three possible error transitions, capable of experimentally suppressing the error to high precision without the need for any calibration, or additional free parameters.
For the two-qubit rotation operator error such as the $ZZ$ error in the effective two qubits' subspace, we present a different approach.
While other schemes typically involve hardware modifications or additional detuned microwave drive terms~\cite{Wei2022Hamiltonian,Mundada2019Suppression,Xu2023ParasiticFree}, we show that a simple, \textsc{drag}-like correction tone applied on the target qubit, along with a detuning on the drive frequency, is sufficient to eliminate dominant entangling error terms while avoiding additional hardware engineering.
An overview of the derived pulse schemes is shown in \cref{fig:overview}c and \cref{fig:overview}d.

In comparison to alternative pulse shaping techniques~\cite{Kirchhoff2018Optimized, Dalgaard2020Global, Baum2021Experimental}, this multi-derivative pulse Ansatz stands out for its simplicity.
It provides an efficient parameterization of the control pulse as a simple expression of the qubits' frequency and anharmonicity.
This simplicity is essential for scalable quantum devices as all the qubits need to be calibrated quickly and repeatedly to ensure high fidelity.
With the qiskit-pulse~\cite{Alexander2020Qiskit} interface, we implemented our drive scheme on multi-qubit devices provided by IBM Quantum.
Despite the limited calibration time due to sporadic access to busy, public machines, our experimental results validate the efficient suppression of coherent errors.
We observe a two- to fourfold reduction in infidelity, achieving beyond state-of-the-art fidelities in the 99.6-99.8\% range on multiple qubit pairs publicly available on the IBM platform.

The rest of the paper is organised as follows:
We start by presenting the theoretical framework of the derivative-based pulse shaping methods.
Next, we derive the pulse schemes for the CR gate and experimentally validate the error suppression for both the transition errors on the control qubit and the multi-qubit errors.
Finally, we demonstrate the performance and scalability of the proposed control scheme by benchmarking the custom-implemented CR gate on multi-qubit quantum hardware, accompanied by numerical simulations across a wide range of experimentally relevant regimes.

\section{Results}
\label{sec: cr drag theory}

\subsection{Multi-derivative pulse shaping}
We start explaining the general theory for the systematic, iterative error suppression with a generic two-level system
\begin{equation}
\label{eq:two-level Hamiltonian}
\hat{H}=\Delta \hat{\Pi}_j+
g(t) \frac{\hat{\sigma}^+_{jk}}{2}
+ g^*(t)\frac{\hat{\sigma}^+_{kj}}{2}
\end{equation}
where $\hat{\Pi}_j=\ket{j}\bra{j}$ and $\hat{\sigma}^+_{jk}=\ket{k}\bra{j}$, $g(t)$ denotes the coupling strength between the two levels.
In the following, we omit the explicit time dependence on $t$ for ease of notation.
In general, $g$ could take the (perturbative) form of an $n$-photon interaction, $\frac{\Omega^n}{\Delta^{n-1}_{\textnormal{eff}}}$, where $\Delta_{\textnormal{eff}}$ is an effective energy gap and $\Omega$ the drive strength.
In particular, if $\Omega$ denotes the CR drive strength on the control qubit, with $n=1$, it describes the transition $\ket{0}\leftrightarrow\ket{1}$ (or $\ket{1}\leftrightarrow\ket{2}$) and with $n=2$ the two-photon transition $\ket{0}\leftrightarrow\ket{2}$.

The goal is to suppress the undesired transition introduced by the coupling $g$.
If $g \ll \Delta$, we may 
perform a perturbative expansion with the antihermitian generator $\hat{S}(\tilde g)=\frac{\tilde{g}}{2\Delta}\hat{\sigma}^+_{jk}-\mathrm{h.c.}$ and obtain under the transformation $\hat{H}'(g)= \hat{V}(\tilde{g})\hat{H}(g)\hat{V}^{\dagger}(\tilde{g})+\im \dot{\hat{V}}(\tilde{g})\hat{V}^{\dagger}(\tilde{g})$ with $ \hat{V}(\tilde{g})=\naturalE^{\hat S(\tilde{g})}$,
\begin{align}
    \hat{H}'(g) = &\im\dot{\hat{S}}(\tilde g) + \hat{H}(g) + [\hat{S}(\tilde g), \hat{H}(g)] + \cdots\\ \nonumber
    = &
    \Delta \hat{\Pi}_j
    +
    \left(
    g
    -
    \tilde{g}
    +
    \im\frac{\dd}{\dd t}
    \frac{\tilde{g}}{\Delta}
    \right) \frac{\hat{\sigma}^+_{jk}}{2}
    +\mathrm{h.c.} + \mathcal{O}(\epsilon^2)
    ,
\end{align}
where $\epsilon \propto g/\Delta$.
We deliberately distinguish between $g$, the actual physical coupling, and $\tilde{g}$, which is used to define the generator $\hat{S}$
that
diagonalizes the Hamiltonian.
As a result, for a time-dependent coupling $g$, to suppress the transition, we require
\begin{equation}
    \label{eq:two-level g solution perturbation}
    g=
    \tilde{g}
    -
    \im\frac{\dd}{\dd t}
    \frac{\tilde{g}}{\Delta}
    .
\end{equation}

The above equation also provides an alternative interpretation:
Transition-less evolution is possible if we find a (counter-diabatic) control $g(t)$ by choosing any continuous function $\tilde{g}(t)$ and making sure that $\hat{S}(\tilde g)$ is zero at the beginning and at the end of the time evolution \cite{Unanyan1997Laserinduced, Demirplak2003Adiabatic,Chen2010Shortcut,Guery-Odelin2019Shortcuts}.
Thus, \cref{eq:two-level g solution perturbation} provides a substitution rule to derive a time-modulated coupling $g(t)$ with the transition between the two levels suppressed.
If the coupling describes an $n$ photon interaction generated by a drive $\Omega$, i.e., $g=\frac{\Omega^n}{\Delta^{n-1}_{\textnormal{eff}}}$ with a constant $\Delta_{\textnormal{eff}}$, we obtain
\begin{equation}
    \label{eq:two-level omega solution perturbation}
    \Omega = 
    \mathcal{F}^{(n)}_\Delta(\tilde{\Omega})
    \coloneqq
    \left(
    {
        \tilde{\Omega}^n - \im\frac{\dd}{\dd t}
        \frac{
        \tilde{\Omega}^{n}
        }
        {\Delta}
    }
    \right)
    ^{\frac{1}{n}}
    .
\end{equation}
Here, we choose $\tilde g=\frac{\tilde \Omega^n}{\Delta^{n-1}_{\textnormal{eff}}}$ to keep the notation intuitive.
The fractional exponent is defined for complex numbers and needs to ensure the continuity of $\Omega$ as a function of $t$.
For $n=1$ this gives the familiar result of the single-derivative \textsc{drag} expansion \cite{Motzoi2013Improving}.
If needed, a free parameter $a$ can be added before the derivative term to adjust the strength DRAG correction.

More generally, a two-level Hamiltonian (or subspace), in \cref{eq:two-level Hamiltonian}, is diagonalized exactly by the unitary transformation (referred to as Givens rotation)~\cite{Li2022Nonperturbative}
\begin{equation}
    \hat{V} = 
    \left(
    \begin{array}{cc}
     \cos \left(\frac{\theta}{2}\right) & \naturalE^{-\im \phi } \sin \left(\frac{\theta }{2}\right) \\
     -\naturalE^{\im \phi } \sin \left(\frac{\theta}{2}\right) & \cos \left(\frac{\theta }{2}\right) \\
    \end{array}
    \right)
    ,
\end{equation}
resulting in an exact substitution rule [c.f.~\cref{eq:two-level g solution perturbation}]:
\begin{equation}
\label{eq:two-level g solution exact}
    g
    =
    \naturalE^{\im \phi } \left(-(\Delta +\dot{\phi})\tan (\theta ) +\im \dot{\theta} \right)
    ,
\end{equation}
where $\theta$ and $\phi$ can in principle be chosen arbitrarily provided $\hat{V}=\mathbbm{1}$ at the beginning and the end of the drive.

This exact diagonalization becomes useful in scenarios involving strong drive amplitudes or small detunings.
In those cases, \cref{eq:two-level g solution exact} provides a compact expression for the DRAG pulse beyond the perturbation limit.
To be consistent with the perturbative solution, we set
$\theta = \arctan(-|\tilde{g}|/\Delta)$ and define $\phi$ as the complex phase of the coupling, i.e., 
$\tilde{g} = \naturalE^{\im \phi} |\tilde{g}|$.
We note that, in general, $\Delta$ could also depend on $g$ and \cref{eq:two-level g solution exact} becomes an implicit equation for $g$ instead of a closed-form expression. To obtain an expression for the drive strength $\Omega$, one needs to invert the dependence of $g=f(\Omega)$.
For instance, for a linear dependence, $g=\kappa \Omega$ (and $\tilde{g}=\kappa \tilde{\Omega}$), with $\kappa$ a constant factor, we get
\begin{equation}
\label{eq:two-level omega 1 solution exact}
    \Omega
    =
     \mathcal{F}^{(1), \textnormal{G}}_\Delta(\tilde{\Omega})
     \coloneqq
     \frac{\Delta +\dot{\phi}_{\tilde{\Omega}}
     }
     {\Delta}
     \tilde{\Omega}
     +
     \frac{\im \naturalE^{\im \phi_{\tilde{\Omega}} }}{\kappa}  \frac{\dd}{\dd t} \arctan(-\frac{|\kappa \tilde{\Omega}|}{\Delta})
\end{equation}
with $\naturalE^{\im \phi_{\tilde{\Omega}}}=\tilde{\Omega}/|\tilde{\Omega}|$.
\Cref{eq:two-level omega solution perturbation} and \cref{eq:two-level omega 1 solution exact} will be the building blocks throughout the remaining of this article as we extend our analysis to multilevel systems.

\subsection{Application to control-qubit errors}
\begin{figure*}[t]
    \includegraphics[]{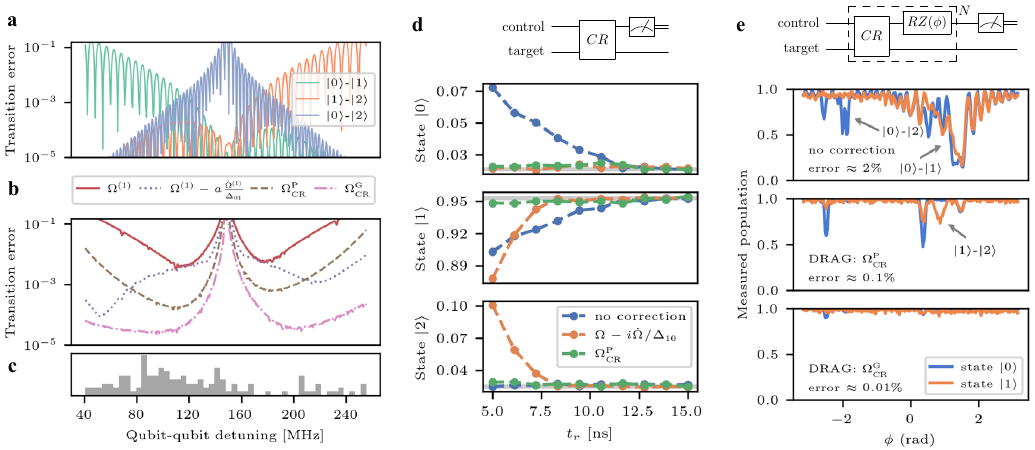}
    \caption{
    The transition errors on the control qubit with different drive schemes. 
    ({\bf a})~Simulated transition error among the 3 levels of the control Transmon introduced by the CR drive using a flat-top Gaussian pulse with $\trise=2\sigma=10$~ns.
    ({\bf b})~The total transition error for different pulse schemes. We plot the envelope of the oscillation by taking the maximum over different pulse lengths with the ramping time $\trise=10$~ns fixed.
    Parameters used are $\Omega_{\textnormal{max}}/2\pi=30$~MHz, $(\Delta_{21}-\Delta_{10})/2\pi=-300$~MHz and $\lambda=\sqrt{2}$.
    ({\bf c})~Distribution of the qubit-qubit detunings in {\it ibm\_brisbane}. A few outliers that are far away out of the studied range are left out.
    ({\bf d})~The circuit and measured transition after preparing the state in $\ket{1}$ and applying a CR pulse with different rising time $\trise$. 
    The data is obtained with a detuning of $120$~MHz and a drive strength of about 40 MHz.
    {\bf e.}~The amplification circuit and measured transition error. Deviations from the expected population of one indicate the presence of error.
    The three plots correspond to a default flat-top Gaussian, recursive \textsc{drag} pulse $\Omega_{\textnormal{CR}}^{\rm{P}}$ in \cref{eq: solution perturbation 3 transitions} and $\Omega_{\textnormal{CR}}^{\rm{G}}$ in \cref{eq: solution exact 3 transitions}, with no calibration of additional parameters.
    The data is obtained from {\it ibm\_nairobi} Q2 and Q1 with the drive amplitude 0.5 ($\approx 60$~MHz), $\trise=10$~ns and $N=30$. The qubit-qubit detuning is about $104$~MHz.
    }
    \label{fig:population error}
    \label{fig:drag tr dependence}
    \label{fig:transition error experiment}
\end{figure*}

The CR interaction is typically activated by driving the control qubit with the frequency of the target~\cite{Magesan2020Effective,Malekakhlagh2020Firstprinciples,Tripathi2019Operation}, leading to a rotation in the target qubit depending on the state of the control, equivalent to a CNOT gate up to single-qubit operations.
Ideally, the state of the control qubit should remain unaltered at the end of the gate.
However, despite the detuning, the drive may still excite the control qubit, especially when operating in the straddling regime for fast entanglement \cite{Tripathi2019Operation, Malekakhlagh2020Firstprinciples}, where the qubit-qubit detuning is smaller than their anharmonicities. Depending on the parameter regimes, it manifests both as single-photon transitions between $\ket{0}\leftrightarrow\ket{1}$, $\ket{1}\leftrightarrow\ket{2}$ as well as the two-photon transition between $\ket{0}\leftrightarrow\ket{2}$~\cite{Malekakhlagh2022Mitigating}.

To counter these transition errors, the single-derivative \textsc{drag} pulse has been employed which introduces a term proportional to the first derivative of the drive pulse, i.e., $\Omega - \im a\frac{\dot{\Omega}}{\Delta}$, with a constant factor $a$ to be optimized~\cite{Malekakhlagh2022Mitigating}.
This heuristic is proven useful when the qubit-qubit detuning is very small, ranging from 50~MHz to 70~MHz~\cite{Kandala2021Demonstration,Wei2024Characterizing}, because in this range there is only one dominant transition.
In contrast, for the scalable, multi-qubit fixed-frequency architecture such as the IBM Quantum plaform, the detuning between neighbouring qubits is distributed over a much broader range~(\cref{fig:population error}c).
An efficient drive scheme must be able to suppress the error in all the operating parameter regimes.
As shown later in this section, in some prevalent parameter regimes, the single-derivative \textsc{drag} pulse provides only minimal improvement.
Even with a numerically optimized scale factor, a compromise arises among different transitions~\cite{Malekakhlagh2022Mitigating}.

Following the multi-derivative pulse described above, we propose the following pulse shape derived by recursively applying the \textsc{drag} correction targeting at the three dominant transitions
\begin{equation}
  \label{eq: solution perturbation 3 transitions}
  \Omega_{\textnormal{CR}}^{\rm{P}}=
  \mathcal{F}^{(1)}_{\Delta_{21}}
  \circ \mathcal{F}^{(1)}_{\Delta_{10}}
  \circ \mathcal{F}^{(2)}_{\Delta_{20}} (\Omega)
\end{equation}
with the perturbative substitution \cref{eq:two-level omega solution perturbation} or
\begin{equation}
    \label{eq: solution exact 3 transitions}
    \Omega_{\textnormal{CR}}^{\rm{G}}=
    \mathcal{F}^{(1),\textnormal{G}}_{\Delta_{21}}
    \circ \mathcal{F}^{(1),\textnormal{G}}_{\Delta_{10}}
    \circ \mathcal{F}^{(2)}_{\Delta_{20}} (\Omega)
\end{equation}
using the exact expression \cref{eq:two-level omega 1 solution exact} for the two single-photon transitions.
The energy difference between state $\ket{j}$ and $\ket{k}$ in the rotating frame is denoted by $\Delta_{jk}$.
The symbol $\circ$ denotes the composition of different substitutions $\mathcal{F}$, applied sequentially from right to left on the pulse shape.
Recursively chaining the \textsc{drag} correction as above suppresses all three dominant errors.
While the two single-photon corrections $\mathcal{F}^{(1)}$ are interchangeable, the substitution for the two-photon transitions needs to be applied first, as detailed in the Methods section.
The explicit formula for the perturbative \textsc{drag} pulse [\cref{eq: solution perturbation 3 transitions}] is given by the following recursive expressions:
\begin{align}
    \Omega_{\textnormal{CR}}^{\rm{P}} &= \Omega_1 - \im \frac{\dot{\Omega}_1}{\Delta_{10}}\\
    \Omega_1 &= \Omega_2 - \im \frac{\dot{\Omega}_2}{\Delta_{21}}\\
    \Omega_2 &= 
    \sqrt{
        \Omega_3^2 - \im \frac{2\Omega_3\dot{\Omega}_3}{\Delta_{20}}
        }
    .
    \label{eq:explicit perturbative recursive DRAG pulse}
\end{align}
Here, $\Omega_3$ needs to be chosen such that the obtained pulse is continuous and starts and ends in zero. Without the last equation for the two-photon transition, the derived pulse aligns with the multi-derivative DRAG solution proposed for multiple linear couplings in \cite{Motzoi2013Improving}.
Notably, if one of the DRAG correction strengths is fine-tuned by an additional parameter, it will not affect the other correction significantly because of the recursive design.

Typically, a CR pulse consists of a rising, a holding and a lowering period, during which the pulse is turned on from zero to the maximum, held for a while and then turned off.
We choose the rising portion of the pulse to be
\begin{align}
\label{eq:pulse shape}
\Omega^{(m)}(t) &= 
\Omega_{\textnormal{max}} \mathcal{I}_0
\int_0^{t} \dd t' \sin^m(\frac{\pi t'}{2\trise}), \quad 0 \le t \le \trise
\end{align}
with the normalization $\mathcal{I}_0$ fixed via $\Omega^{(m)}(\trise)=\Omega_{\max}$.
This definition ensures that the pulse is $m$ times differentiable and the derivatives are zero at $t=0$ and $t=\trise$, which guarantees the validity of the frame transformation $\hat{V}$ introduced above.
Other pulse shapes can also be used as long as this property is satisfied.
After the holding time, the lowering phase takes the time-reversed shape.
An example of the CR pulse is shown in \cref{fig:overview}c.
For $m=1$ and with zero holding time, the pulse is the same as the Hann window, very close to the flat-top Gaussian pulse commonly adopted.
It is important to note that, as $m$ increases, more high-frequency components become incorporated into the pulse shape, leading to higher non-adiabatic transition error if not compensated for.
Therefore, it is advisable to keep $m$ as small as possible.
For our study, we use $m=3$ as the initial shape for the recursive \textsc{drag} pulse.

To verify the performance of the error suppression, we numerically simulate the dynamics of the three-level Hamiltonian of the control Transmon (see Methods) and the result is shown in \cref{fig:population error}a and \cref{fig:population error}b.
First, we examine in \cref{fig:population error}a the contribution of the three transition errors for an uncorrected pulse, across the typical experimentally relevant qubit-qubit detuning values.
The error is defined as the probability of unwanted population transfer among different states.
The plot indicates that all three transitions must be considered for sufficient error suppression.
Moreover, we observe that partial suppression of the errors (using only one or two derivatives) may increase the unsuppressed ones, as demonstrated in detail in Supplementary Note 1, making them non-negligible even if they were initially small.

Next, we compare the total transition error introduced by different pulse schemes in ~\cref{fig:population error}b.
To better illustrate the difference between the pulse schemes, we take the sum of the three transition errors and the maximum over pulses with various holding lengths.
In this way, the oscillation caused by the pulse timing is removed and only the upper envelope remains.
As a baseline, we plot the error for pulse shape $\Omega^{(1)}$, which is similar to the flat-top Gaussian pulse used in qiskit-pulse~\cite{Alexander2020Qiskit}.
The recursive \textsc{drag} pulse shapes we derived, $\Omega_{\textnormal{CR}}^{\rm{P}}$ and $\Omega_{\textnormal{CR}}^{\rm{G}}$, suppress the error by several orders of magnitude, without any numerical optimization, as long as the drive is not resonant with the two-photon transition.
In comparison, the single-derivative \textsc{drag} pulse, used in previous works~\cite{Kandala2021Demonstration,Wei2024Characterizing,Malekakhlagh2022Mitigating}, performs well only when the error is dominated by one single-photon transition (very large or very small detuning).
Outside of these regimes, its performance is restricted due to the compromise between different transitions, even if the \textsc{drag} coefficient $a$ is calibrated to minimize the total error.

This observation is further supported by experimental results shown in \cref{fig:drag tr dependence}d and \cref{fig:transition error experiment}e.
In \cref{fig:drag tr dependence}d, the state was initialized in state $\ket{1}$, and a CR pulse of 200 ns with varying rising times $\trise$ was applied.
As the rising time decreases, the error transition grows quadratically.
Without any correction, the error is dominated by the transition between $\ket{0}\leftrightarrow\ket{1}$.
Applying a \textsc{drag} pulse designed to suppress this transition, i.e., $\Omega-\im \dot{\Omega}/\Delta_{10}$, effectively suppresses this error but introduces a new transition error between $\ket{1}\leftrightarrow\ket{2}$.
Calibrating the \textsc{drag} coefficient only compromises between these errors.
In contrast, with the recursive pulse shape defined in \cref{eq: solution perturbation 3 transitions}, all errors are suppressed below the state preparation and measurement error.

Typically, achieving high-fidelity quantum operations requires the transition errors to be suppressed to the order of $10^{-4}$.
Resolving this population error often needs a large number of sampling points.
Therefore, we employ the error amplification circuits outlined in~\cite{Wei2024Characterizing},
which add virtual Z gates $RZ(\phi)$ between the repetitions with different phases $\phi$ (\cref{fig:transition error experiment}).
Different transition errors will be selected by different choices of $\phi$, as detailed in Supplementary Note 3.
To perform the measurement, we calibrate an X gate between states $\ket{1}$, $\ket{2}$ and build a measurement discriminator for qutrits~\cite{Morvan2021Qutrit}.

In \cref{fig:transition error experiment}e, the measured population of the state $\ket{0}$($\ket{1}$) is plotted after the initial preparation in $\ket{0}$($\ket{1}$) and 30 repetition of the CR pulse. A population close to one implies negligible errors, while any deviation indicates a transition to other states. For instance, overlapping blue and orange curves indicate the $\ket{0}\leftrightarrow\ket{1}$ transition, while a drop solely in the blue curve suggests the $\ket{0}\leftrightarrow\ket{2}$ transition.
It is evident that for this short rising time ($\trise=10$~ns), there exists a significant transition error between state $\ket{0}$ and $\ket{1}$, but also a non-negligible contribution from other transitions between $\ket{0}$ and $\ket{2}$.
After applying the perturbative \textsc{drag} pulse, a substantial reduction in the error is observed, with some remaining small transitions.
Using the recursive \textsc{drag} pulse derived by Givens rotation proves highly effective, suppressing all transition errors below the threshold.
In both cases, no calibration of \textsc{drag} coefficients is required, and the analytical formulas are completely predictive.
In general, free parameters can be added to each substitution before the derivative terms to fine-tune the strength of DRAG corrections for each individual transition error.

It is crucial to highlight that previous applications of a single-derivative \textsc{drag}~\cite{Kandala2021Demonstration,Wei2024Characterizing,Malekakhlagh2022Mitigating} to CR gates primarily focus on the case of very small qubit-qubit detuning (ranging from 50 MHz to 70 MHz), the errors of which is dominated only by the $\ket{0}\leftrightarrow\ket{1}$ transition.
However, qubit pairs on IBM Quantum Platform have detuning distributed in a much larger range from 40 to 260 ~MHz (\cref{fig:population error}c), where other transitions become nonnegligible~(\cref{fig:population error}a).
In contrast, the recursive \textsc{drag} solution showcased in this study exhibits remarkable universal performance even in the presence of multiple types of errors, without any calibration necessary.
In Supplementary Note 5, we show similar error suppression on qubit pairs with qubit-qubit deutning of 143~MHz and 189~MHz, together with an example where the single-derivative \textsc{drag} fails to suppress the error even with a full-sweep calibration of the \textsc{drag} coefficient.

\subsection{Application to multi-qubit operator errors}
\label{sec: suppressing dynamic ZZ}
\begin{figure}[t]
\centering
\includegraphics[]{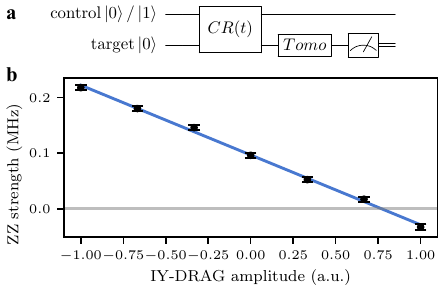}
\caption{
Measurement of the ZZ error and calibration of the IY-DRAG schemes.
({\bf a})~Circuit used for the CR Hamiltonian tomography~\cite{Sheldon2016Procedure}.
({\bf b})~Experimental data for calibrating the $IY$-\textsc{drag} amplitude to minimize the $ZZ$ coupling, obtained from {\it ibm\_perth}. The error bar denotes the standard deviation of the measured $ZZ$ strength.
}
\label{fig:ZZ IZ correction}
\end{figure}

A second major part of the error in the CR operation comes from the remaining dynamical operators in the two-qubit subspace that do not commute with the ideal dynamics $ZX$.
Assuming the transition errors on the control qubit are all suppressed, the effective Hamiltonian in the two-qubit subspace is given by
\begin{align}
    H =
    &\frac{\nu_{ZX}}{2} \hat{Z}\hat{X} +
    \frac{\nu_{ZY}}{2} \hat{Z}\hat{Y} +
    \frac{\nu_{ZZ}}{2} \hat{Z}\hat{Z} \nonumber \\+
    &\frac{\nu_{IX}}{2} \hat{I}\hat{X} +
    \frac{\nu_{IY}}{2} \hat{I}\hat{Y} +
    \frac{\nu_{IZ}}{2} \hat{I}\hat{Z}
    .
    \label{eq:effective hamiltonian cr}
\end{align}
The coefficient $\nu(t)$ for each term can be derived by perturbative expansion, with the explicit expressions given in appendix C of Ref.~\cite{Magesan2020Effective}.
Experimentally, they can be measured by Hamiltonian tomography~\cite{Sheldon2016Procedure}.
An overview of the multi-qubit errors and the frame transformations used below to remove them is shown in \cref{fig:overview}d.

When implementing a CR gate, the $ZY$ term is removed by calibrating the phase of the CR drive and the single-qubit rotations, $IX$ and $IY$, compensated for by a target drive~\cite{Sheldon2016Procedure}.
To achieve high-fidelity operations, we iteratively fine-tune the drive pulse until the error terms $ZY$, $IY$ and $IX$ are all below 0.015~MHz (see Supplementary Note 6).
After this standard calibration procedure, one can describe the dynamics with the following effective Hamiltonian
\begin{equation}
    \hat{H} = \frac{\nu_{ZX}}{2} \hat{Z}\hat{X} + \frac{\nu_{ZZ}}{2} \hat{Z}\hat{Z} + \frac{\nu_{IZ}}{2} \hat{I}\hat{Z}
    ,
    \label{eq:effective hamiltonian cr after calibration}
\end{equation}
where the first term is the desired Hamiltonian dynamic while the other two are multi-qubit errors to be suppressed.

We now show that an $IY$-\textsc{drag} correction and a detuning are sufficient to suppress the remaining errors.
Note that the two Hamiltonian terms $ZX$ and $ZZ$ are connected by a rotation along the $IY$ axis.
Hence, we define the transformation
\begin{equation}
    \hat{V}_{ZZ} = \hat{I} \otimes \exp(-\im\beta(t) \hat{Y}/2)
\end{equation}
with
$\beta = \arctan(\frac{\nu_{ZZ}}{\nu_{ZX}})\approx \frac{\nu_{ZZ}}{\nu_{ZX}}$.
In the perturbaitve expansion, the coefficients are given as $\nu_{ZZ} = \frac{J^2|\Omega_{\rm{CR}}|^2}{2\Delta_{\rm{eff}}(\alpha_1 + \Delta)2}$ and $\nu_{ZX} + \im \nu_{ZY}= - \frac{J\Omega_{\rm{CR}}\alpha_1}{\Delta (\alpha_1 + \Delta )}$, where $\Delta_{\rm{eff}}$ is a constant depending on the Transmons' frequency and anharmonicity~\cite{Magesan2020Effective}.
This transformation results in an enhanced $ZX$ strength $\sqrt{\nu_{ZX}^2 + \nu_{ZZ}'^2}$ and an additional single-qubit term $\dot{\beta} \hat{I}\hat{Y}/2$ to be compensated by a $\hat{Y}$ drive on the target qubit.
It is not difficult to verify that this corresponds to a \textsc{drag}-like correction which is non-zero only during the pulse ramping time.

In general, to completely remove the error, one needs to match the shape of the $IY$-\textsc{drag} pulse exactly with $\dot{\beta}/2$.
In typical CR gates, the holding period is much longer than the ramping time.
Therefore, we can neglect the coherent error introduced by the time-dependent part $0<t<\trise$ and focus only on the holding period $\trise<t<\tfinal-\trise$. This simplified approach allows us to neglect the shape of the $IY$-\textsc{drag} pulse and only calibrate the area (amplitude) such that the $ZZ$ error is removed during the holding period.
We choose the $IY$-\textsc{drag} shape as the first derivative of the target drive, i.e., $c_{IY}\dot{\Omega}_{IX}$.
Given that the $ZZ$ error is typically small (<0.1~MHz), the $IY$-\textsc{drag} correction is also very weak.
Thus, the correct coefficient $c_{IY}$ can be obtained by measuring the $ZZ$ coupling strength for a few different $c_{IY}$ and conducting a linear fit, as illustrated in \cref{fig:ZZ IZ correction}.
In practice, we find that three sampling points are sufficient for the accurate calibration of the $IY$-\textsc{drag} amplitude.
It is worth noting that in the calibration, the removed $ZZ$ error consists of both the dynamic ones introduced by the drive and the static $ZZ$ terms caused by residual coupling~\cite{Xu2021ZZ}.

Compared to the previous approach applied by IBM, the target rotatory pulse~\cite{Sundaresan2020Reducing}, our proposed method require only three sampling points and a linear fit, employing the same tomography circuit as used in the standard calibration~\cite{Sheldon2016Procedure}, which renders it more practical for implementation on the IBM platform with limited calibration time.
In contrast, the calibration of the target rotatory pulse amplitude requires a sweep across various amplitudes and finding a minimum of total measured errors.
Furthermore, our method does not require the echoed CNOT structure and thus can be used to construct a direct CNOT gate.
The two methods can also be combined, introducing new degrees of freedom to suppress more residual errors at the same time, which are left for future study.

Finally, the only untreated error, the $IZ$ term, is compensated for by detuning the CR drive.
In general, the exact cancellation of the $IZ$ error requires time-dependent detuning, i.e., a chirped pulse or phase ramping.
Here, as the $IZ$ term is usually small, it is sufficient to compensate for it with a constant detuning.
This is implemented by adding an additional phase term to the pulse shape $\exp(-\im\nu_{IZ} t/2)$, where $\nu_{IZ}$ denotes the measured $IZ$ coefficient from the tomography, similar to the phase error in single qubit gates~\cite{Chen2016Measuring}.

\subsection{Benchmarking the improved CR gate}

\begin{figure}[t]
\centering
\includegraphics[width=0.95\linewidth]{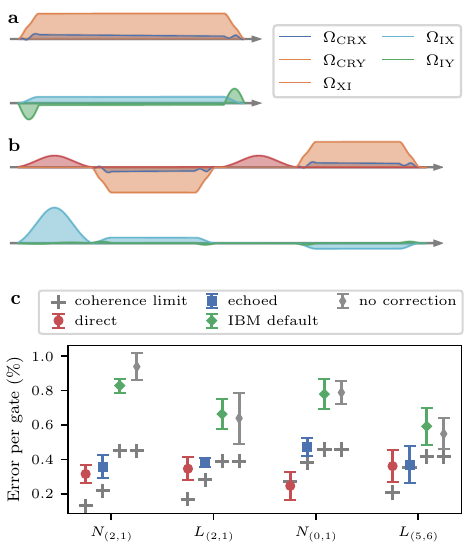}
\caption{
Calibrated pulses for the CNOT gate and randomized benchmarking.
Examples of the custom pulse shapes used are shown for the direct ({\bf a}) and echoed ({\bf b}) CNOT gate on {\it ibm\_lagos} qubit Q2 and Q1.
The amplitudes for different drives are rescaled for visualization purposes.
({\bf c}) Error per gate of the self-calibrated CNOT and the default CNOT gate, estimated by interleaved randomized benchmarking. The coherence limit is calculated including relaxation (T1) and dephasing (T2) noise provided by IBM Quantum. Data obtained from {\it ibm\_lagos} ($L$) and {\it ibm\_nairobi} ($N$).
The error bar represents the standard deviation of five randomized benchmarking experiments.
}
\label{fig:exp fidelity}
\end{figure}

The investigations outlined above underscore the performance of our proposed methods in addressing both the transition errors on the control Transmon induced by rapid driving and the multi-qubit operator errors arising from (static and dynamic) residual coupling.
The improved precision in control not only reduces the coherent error but also facilitates the exploration of higher drive amplitudes and faster tuning speeds, which usually introduces more coherent error if left uncompensated~\cite{Tripathi2019Operation,Malekakhlagh2020Firstprinciples}.
As a result, the attained reduction in gate time allows us to exceed the impact of decoherence and achieve higher fidelities.
For instance, by reducing $\trise$ from 28~ns (default qiskit-pulse parameter) to 10~ns, one gains about 35~ns for an echoed CR gate.
Moreover, because of the simplified calibration procedure, a complete removal of multi-qubit errors is achieved only when the drive is at its maximum.
Therefore, the reduction in $\trise$ contributes not only to shorter gate time but also to improved accuracy in the CR drive.

In consideration of these factors, we select pairs of qubits from the available Transmon qubits on the IBM Quantum Platform that exhibit sufficiently large coupling and relatively long coherence time (>30$\mu$s).
Due to our limited access, a comprehensive search for the speed limit at different drive strengths to determine the optimal gate time is not feasible.
Therefore, we opt for an empirical approach in choosing the CR drive amplitude based on the qubit-qubit detuning and the effective coupling strength.
In addition to the commonly used echoed CNOT gates~\cite{Sheldon2016Procedure}, we also calibrate direct CNOT gates, eliminating the two single-qubit gates and further reducing the gate time.
Examples of the applied pulse shapes are shown in \cref{fig:exp fidelity}a and b, which include the recursive-\textsc{drag} pulse correction on the CR drive (ramping up of the CR pulse), the $IY$-\textsc{drag} correction (green pulse), and the frequency detuning (asymmetry between the ramping up and off).
The calibration procedure is explained in detail in Supplementary Note 6.

To accurately characterize the fidelity, we measure the infidelity of the self-calibrated CR gates through interleaved randomized benchmarking.
For each pulse configuration, we repeat the experiment five times (each takes about 5 to 10 minutes including the classical communication time) and compute the mean and standard deviation of the measured gate error.
Therefore, the presented gate error should be interpreted as the average error over the following hours after the calibration, including possible detrimental drift in the prior system parameters over the ensuing time period.
As shown in \cref{fig:exp fidelity}, we obtain a significant reduction of the error on several pairs of the qubits compared to the default CNOT gate. 
Over the four pairs of qubits studied, we obtain an average gate fidelity of 99.7(1)\%.
We further compare it to the self-calibrated CNOT gate with no corrections applied, using the default Gaussian shape, drive amplitude and $\trise$.
This comparison verifies that the observed improvement is not solely attributable to our more recent calibration.
Detailed information on the used qubits, measured effective coupling strength and drive parameters is presented in Supplementary Note 4.

\begin{figure*}[t]
    \centering
    \includegraphics[width=\textwidth]{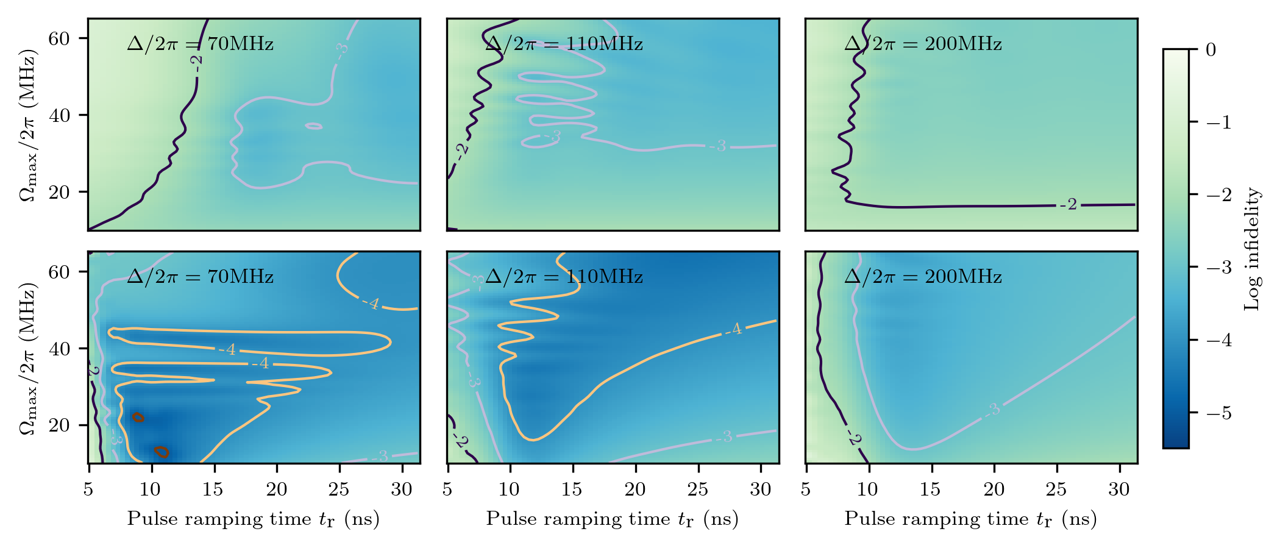}
    \caption{
    Infidelities of simulated CR gates implemented by the flat-top Gaussian pulse (top) and the proposed multi-derivative pulse (bottom), including the derived recursive \textsc{drag} correction ($\Omega_{\rm{CR}}^{\rm{G}}$) and the $IY$-\textsc{drag} correction. All the pulses are closed-form and computed deterministically without numerical calibration. The simulation is repeated for 3 different values of the qubit-qubit detuning $\Delta$.
    }
    \label{fig:improvement_infidelity}
\end{figure*}

Next, we characterize the possible improvement over a wider range of typical and prospective parameter regimes. In particular, we perform thorough numerical simulations to demonstrate what is possible beyond current bottlenecks given by present-day coherence times and limited calibration access on high-demand systems.

To show the applicability of the derived pulse on a large-scale quantum device, it is important to evaluate its performance across various parameter regimes representative of a real quantum system.
The Hamiltonian model is chosen to have similar coupling strength and $ZZ$ error rate to those qubits on the IBM Quantum Platform (see Methods).
Since we focus on the CR operation, we exclude the error that can be removed by single-qubit corrections.
We perform a sweep for different $\Omega_{\max}$ and $\trise$ while also varying the qubit-qubit detuning $\Delta/2\pi$=$\{70,110,200\}$~MHz in the straddling regime.

In \cref{fig:improvement_infidelity}, we compare the infidelity between the proposed pulse and the flat-top Gaussian pulse.
It shows a drastic reduction in the coherent error in all regimes via our approach, with orders of magnitude suppression similarly seen for the three detunings.
The recursive \textsc{drag} correction reduces the ramping time while keeping a low transition error rate.
Meanwhile, the $IY$-\textsc{drag} correction cancels the $ZZ$ error and allows for stronger drive amplitude.
The observed optimal selection of the pulse ramping time between 10 to 15 ns in the simulations results from a compromise between the static $ZZ$ error in IBM qubit parameters and the transition error. 
In our simulation, we considered IBM hardware with fixed coupler frequencies, resulting in a static $ZZ$ error that cannot be fully corrected, especially during the ramping period. 
Shorter ramping times lead to reduced accumulation of this static $ZZ$ error, at the expense of increased transition error.

For commensurate qubit lifetimes in the range of milliseconds, as already demonstrated in Refs.~\cite{Somoroff2023Millisecond,Wang2022Practical}, errors as low as $10^{-4}$ are within reach using our proposed pulse, while the standard pulse would be limited to an order of magnitude larger error.
Along with the shorter gate times coming from larger amplitudes and shorter ramping times, the large coherent error suppression further amplifies any expected gains coming from improvements in qubit fabrication. This is already seen, for example, in Fig.~\ref{fig:exp fidelity}, where not only is the coherent error on the IBM Quantum devices suppressed, but there is also a reduction in the total gate duration (with reduced coherence limit).  This is especially important looking forward, as advantages in coherence times for fixed-frequency architectures vs. tunable-qubit architectures tilts the advantage towards the former with appropriate pulse shaping.
Note also that even if the parasitic $ZZ$ error is engineered to be very small~\cite{Goerz2017Charting,Li2022Nonperturbative, Mundada2019Suppression,Xu2023ParasiticFree}, as coherence times improve, the standard pulses must choose a long ramping time to match the incoherent error, while our pulse shaping approach can continue to use very short times, fully taking advantage of such improvements.

Importantly, these pulses are constructed following the analytical expression without additional optimization or fitting parameters. This means that compared to all but the simplest approaches available, including Ref.~\cite{Malekakhlagh2022Mitigating}, these high-performance pulses are much faster and more straightforward to calibrate. Additionally, we observe that the transition error is barely affected by the drift of the drive strength and is also relatively robust against frequency drift (see Supplementary Note 2)

\section{Discussion}
We introduced an analytical multi-derivative pulse shape tailored for driving the CR interaction in superconducting qubits, adept at eliminating undesired transitions on the control qubit and unwanted multi-qubit dynamics.
Our approach extends the \textsc{drag} formalism to a recursive structure capable of suppressing multiple error transitions simultaneously.
Additionally, we developed a novel technique to eliminate multi-operator errors by dynamically transforming the errors into the desired entangling form. 
This resulted in several orders of magnitude suppression in the coherent error when simulating across the range of typical c-QED regimes, without extensive requirement on calibration.
The simplicity and universality of the proposed pulse shape make it well-suited for implementation on the IBM Quantum Platform as an efficient high-quality calibration across hundreds of qubits.
We demonstrate this on several qubits, showing a significant suppression in the state-of-the-art error using our customized pulse shape. The results are reproducible over a wide range of qubit frequency spacings and with prescriptive pulse shapes across the spectrum.

These analytical approaches are general and also applicable to other entangling gates in c-QED and various quantum technologies~\cite{Nesterov2022CNOT,Dogan2023TwoFluxonium,Ficheux2021Fast}.
The control error addressed in this work extends beyond CR gates and is relevant to other off-resonant drive schemes, such as microwave-activated gates~\cite{Mitchell2021HardwareEfficient,Wei2022Hamiltonian,Kim2022Highfidelity,Goss2022Highfidelity}, as well as the use of microwave drives for suppressing quantum cross-talk and leakage~\cite{Motzoi2013Improving,Zhao2022Quantum,Cai2021Impact,Marques2023Allmicrowave}. The coherent error suppression demonstrated here also has implications for fixed-frequency architectures, allowing them to take advantage of longer coherence times compared to tunable architectures.
Moreover, errors involving a spectator qubit~\cite{Zhao2022Quantum,Cai2021Impact,Malekakhlagh2020Firstprinciples, Sundaresan2020Reducing} can be addressed by incorporating the ancillary level into the modeling and introducing new derivative-based corrections accordingly.

Apart from the pursuit of improving multi-qubit gates, it is noteworthy that the suppression of coherent errors also indirectly enhances the fabrication process' yield.
For instance, the transition errors addressed in this work were also identified as frequency collisions in Ref.~\cite{Zhang2022Highperformance,Hertzberg2021Laserannealing}.
The proposed drive scheme effectively increases the threshold for frequency collisions, thereby contributing to an increased fabrication yield.
Similar error models for frequency collision also apply to the tunable-coupler architecture~\cite{Osman2023Mitigation}, extending the potential application domain.

\section{Methods}

\subsection{Derivation of the recursive \textsc{drag} pulse}
\label{sec: Three-step derivation}

We use the following three-level Hamiltonian to model the control Transmon
\begin{equation}
    \label{eq:3-level model}
    \hat{H} = \frac{\epsilon\Omega_{
    \rm{CR}}}{2}(\sigma^+_{01} + \lambda \sigma^+_{12}) + \mathrm{h.c.} + \Delta_{10} \hat{\Pi}_1 + (\Delta_{10}+\Delta_{21}) \hat{\Pi}_2
    ,
\end{equation}
where $\lambda$ is the relative coupling strength of the second transition and $\epsilon$ is used to denote the perturbation order.
For detuning $\Delta_{10}=0$, the pulse is on resonance and implements a single-qubit gate.
When the drive is resonant with the frequency of the target qubit, a CR operation is activated.
In the rotating frame with respect to the driving frequency, we have $\Delta_{10}$ equal to the qubit-qubit detuning and $\Delta_{21}=\Delta_{10}+\alpha_{\rm{c}}$, with $\alpha_{\rm{c}}$ the anharmonicity.
To the leading order perturbation, the coupling strength is proportional to $\Omega_{\rm{CR}}$~\cite{Magesan2020Effective}.
An ideal CR pulse generates rotations on the target qubit depending on the control qubit state while leaving the latter intact.
This approximation holds well as long as the dressing of the qubit is perturbative.
Therefore, we aim at finding a pulse $\Omega_{\rm{CR}}$ with non-zero real integral but introducing no population transfer among any of the three levels of the control qubit.
This model, \cref{eq:3-level model}, includes both the leakage error and population flipping on the control qubit~\cite{Tripathi2019Operation}.

In the following, we show the derivation of the substitution rule of $\Omega$ in \cref{eq: solution perturbation 3 transitions} via Schrieffer Wolff perturbation.
We omit the perturbative corrections to the diagonal part of the Hamiltonian as they have no effect on the leading-order perturbative coupling strength.
The derivation includes three steps, each targeting one coupling.
The perturbative transformation generated by an anti-hermitian matrix $\hat{S}$ is defined as
\begin{equation}
\hat{H}' = \im\dot{\hat{S}} + \hat{H} +\left[\hat{S},\hat{H}\right]+\frac{1}{2} \left[\hat{S},\left[\hat{S},\hat{H}\right]\right] + \cdots
.
\end{equation}

First, we apply the perturbative diagonalization targeting the $\ket{0}\leftrightarrow\ket{1}$ transition
\begin{equation}
\label{eq:substitution S1}
\hat{S}_1 = \frac{\epsilon}{2} \left(\frac{\Omega_1 }{\Delta _{10}}\hat{\sigma}^+_{01} + \frac{\lambda \Omega_1 }{\Delta _{10}}\hat{\sigma}^+_{12}\right) - \mathrm{h.c.}
\end{equation}
The first component in $\hat{S}_1$ is chosen to remove the $\ket{0}\leftrightarrow\ket{1}$ coupling perturbatively.
According to the derivation in the main text, we define a substitution for $\Omega_{\rm{CR}}^{\rm{P}}$
\begin{equation}
   \Omega_{\rm{CR}}^{\rm{P}} = \Omega_1 - \im \frac{\dot{\Omega}_1}{\Delta_{10}}
    .
\end{equation}
The second term in \cref{eq:substitution S1} is chosen such that $\im \dot{\hat{S}}_1$ is proportional to the $Y$ control Hamiltonian.
This ensures that in the derived effective Hamiltonian, no $\dot{\Omega}_1$ appears in the $\ket{1}\leftrightarrow\ket{2}$ coupling, because it is absorbed in $\Omega_{\rm{CR}}$.
Note that it does not diagonalize the $\ket{1}\leftrightarrow\ket{2}$ coupling, which would need $\frac{\lambda  \Omega_1 }{\Delta _{21}}\hat{\sigma}^+_{12}$ instead.
As a result, we obtain
\begin{align}
\hat{H}_1 &= 
    \left(1 -\frac{\Delta _{21} }{\Delta _{10}}\right)
    \left(
    \frac{1}{2} \lambda  \Omega _1 \epsilon
    \hat{\sigma}^+_{12}
     -
     \frac{\lambda  \Omega _1^2 \epsilon ^2}{8 \Delta _{10}}
     \hat{\sigma}^+_{02}
     \right)
     \nonumber
     +
     \mathrm{h.c.}\\
     & +  \textnormal{diag} +\mathcal{O}(\epsilon^3)
     .
\end{align}

In the second step, we perform another perturbative diagonalization that removes the $\ket{1}\leftrightarrow\ket{2}$ transition:
\begin{equation}
S_2 = 
\left(1 -\frac{\Delta _{21} }{\Delta _{10}}\right)
\frac{ \lambda  \Omega _2 \epsilon}{2 \Delta_{21}}
\hat{\sigma}^+_{12}
 -  \mathrm{h.c.}
\end{equation}
and substitute 
\begin{equation}
    \Omega_1 = \Omega_2 - \im \frac{\dot{\Omega}_2}{\Delta_{21}}
    .
\end{equation}
This gives the effective Hamiltonian
\begin{align}
\label{eq:H2}
\hat{H}_2 &= 
    \left(
    \frac{\Delta _{21}}{\Delta _{10}}-1
    \right)
    \left(
    \Omega_2 - \im \frac{\dot{\Omega}_2}{\Delta_{21}}
    \right)^2
    \frac{\lambda   \epsilon ^2}{8 \Delta _{10}}
    \hat{\sigma}^+_{02}
      +\mathrm{h.c.}\\ \nonumber
      &+  \textnormal{diag} +\mathcal{O}(\epsilon^3)
\end{align}
where both single-photon transitions are removed to the leading order.
Notice that the \textsc{drag} pulse shape is independent of the relative drive amplitude $\lambda$ in this first-order approximation.

It may seem strange that the remaining coupling for the $\ket{0}\leftrightarrow\ket{2}$ transition is not symmetric with respect to the order of the transformations of $\ket{0}\leftrightarrow\ket{1}$ and $\ket{1}\leftrightarrow\ket{2}$, although the two substitutions commute.
In fact, we can perform a transformation $\hat{S}_3$
\begin{equation}
    \hat{S}_3 = 
    -\left(\frac{1}{\Delta _{21}}-\frac{1}{\Delta _{10}}\right)
    \frac{ \left(\lambda  \Omega _2^2 \epsilon ^2\right)}{8 \Delta _{10}}
    \hat{\sigma}^+_{02}
    - \mathrm{h.c.}
\end{equation}
which only removes the $\Omega \dot{\Omega}$ term and gives
\begin{align}
H_3 &= 
    \frac{1}{8}\lambda
    \epsilon ^2
    \left(
        \frac{1}{\Delta _{21}}-\frac{1}{\Delta _{10}}
    \right)
    \left(
    \frac{\dot{\Omega}_2^2}{\Delta _{21}\Delta _{10}}
    +
    \Omega_2^2
    \right) 
    \hat{\sigma}^+_{02}
    +\mathrm{h.c.}\\\nonumber
    &+  \textnormal{diag} +\mathcal{O}(\epsilon^3)
    .
\end{align}

Lastly, we perform the third step to suppress the remaining $\ket{0}\leftrightarrow\ket{2}$ coupling.
To fully remove this transition one needs to solve the differential equation
\begin{flalign}
    \left(
    \frac{\dot{\Omega}_2^2}{\Delta _{21}\Delta _{10}}
    +
    \Omega_2^2
    \right)
    = \\\nonumber
    \left(
    \frac{\dot{\Omega}_3^2}{\Delta _{21}\Delta _{10}}
    +
    \Omega_3^2
    \right) 
    &-
    \im \frac{
    \dd
    }{\dd t}
    \frac{
    \left(
    \frac{\dot{\Omega}_3^2}{\Delta _{21}\Delta _{10}}
    +
    \Omega_3^2
    \right) 
    }
    {\Delta_{20}}
    ,
\end{flalign}
which is difficult because of the non-linearity.
Moreover, it may result in a pulse that does not fulfil the boundary condition, unless $\Omega_3$ is carefully chosen to ensure that.
In practice, numerical solutions may be employed to solve the equation, though it will pose challenges for fast calibration.
For simplicity, we here assume that the pulse ramping is quasi-adiabatic i.e. $\Omega_2^2 \gg \frac{\dot{\Omega}_2^2}{\Delta_{10}\Delta_{21}}$.
For the parameters studied in this work, with $\dot{\Omega} \approx \frac{\Omega}{\trise}$, this threshold lies around $t_r
\approx 6$~ns.
In this case, we can ignore the term proportional to $\dot{\Omega}^2_2$.

We then define the last transformation that diagonalizes the $\ket{0}\leftrightarrow\ket{2}$ transition
\begin{equation}
\hat{S}_4 = 
    \frac{1}{8}\lambda
    \epsilon ^2
    \left(
        \frac{1}{\Delta _{21}}-\frac{1}{\Delta _{10}}
    \right)
    \frac{\Omega_2^2}{\Delta_{20}}
    \hat{\sigma}^+_{02}
    -\mathrm{h.c.}
\end{equation}
and substitute
\begin{equation}
    \Omega_2 = 
    \sqrt{
        \Omega_3^2 - \im \frac{2\Omega_3\dot{\Omega}_3}{\Delta_{20}}
        }
        .
\end{equation}
Here, $\Omega_3 = \Omega^{(3)}$, defined in \cref{eq:pulse shape}, which is a continuously three-times differentiable function and ensures that the final expression starts and ends at zero.
As a result, we suppress all three transitions up to $\mathcal{O}(\epsilon^3) + \mathcal{O}(\dot{\Omega}^2/\Delta^4)$.

Combining the three expressions, we obtain the explicit formula for the perturbative recursive \textsc{drag} pulse presented in the main text.
As simple as the perturbative \textsc{drag} expression is, 
it may not sufficiently suppress the error if the qubits frequencies are very close to the one of $\ket{0}\leftrightarrow\ket{1}$ or $\ket{1}\leftrightarrow\ket{2}$ and the perturbative approximation no longer holds, as shown in~\cref{fig:population error}b.
To address this limitation, we replace the substitutions for the single-photon transitions with the exact diagonalization based on Givens rotations defined in \cref{eq:two-level omega 1 solution exact}.
It is important to note that the substitution $\mathcal{F}^{(1),G}$ is only exact concerning the two-level subsystem; corrections to the energy gaps and other couplings are still disregarded.
Nevertheless, it still significantly improves the performance compared to the perturbative expressions.

\subsection{Numerical simulation of the CR gate}
In the simulation, we use an effective Duffing model \cite{khani2009optimal} truncated at 4 levels
\begin{align}
    \hat{H}_0 = \omega_a \hat{a}^{\dagger} \hat{a} + \sum_{j=1,2}\omega_j \hat{b}_j^{\dagger} \hat{b}_j + \frac{\alpha_j}{2} \hat{b}_j^{\dagger} \hat{b}_j^{\dagger} \hat{b}_j\hat{b}_j
    + g_j (\hat{b}_j \hat{a}^{\dagger} + \hat{b}_j^{\dagger} \hat{a})
\end{align}
where $\hat{b}_j$ and $\hat{a}$ are the annihilation operators for qubit $j$ and the resonator, respectively, and $g_j$ is the coupling strength.
The microwave drive on qubit $j$ is written as
\begin{align}
    \hat{H}_{\rm{c}} &= 
    \Re(\Omega_{\rm{CR}})\cos(\omega_{\textnormal{d}} t) (\hat{b}_j^{\dagger} + \hat{b}_j) \\\nonumber
    &+
     \im \Im(\Omega_{\rm{CR}})\cos(\omega_{\textnormal{d}} t) (\hat{b}_j^{\dagger} - \hat{b}_j)
\end{align}
where $\omega_{\textnormal{d}}$ is the driving frequency, initially chosen as the frequency of the target qubit.
For simplicity, we use the same drive frequency for both the control and the target qubit.

We choose the anharmonicity $\alpha=-300$~MHz and $g_j=80$ MHz.
The detuning of the coupler from the control qubit, i.e., $\omega_1-\omega_r$, is about -1.4 GHz and adjusted such that the effective qubit-qubit coupling strength is about 3~MHz, with $ZZ$ crosstalk around 0.06~ MHz, similar to the Transmons on the IBM platform (see Supplementary Note 4).
Based on the model above, we derive the CR pulse following the analytical expressions derived in this paper.
The effective coupling strength of $ZX$ and $ZZ$ are computed using the Non-Perturbative Analytical Diagonalization (\textsc{npad}) method~\cite{Li2022Nonperturbative}, from which the gate time, i.e., the holding duration of the pulse, is calculated.

When computing the fidelity in the simulation, we ignore the contribution of the (commuting) single-qubit corrections $\hat{Z}\hat{I}$ and $IX$, because they can be easily calibrated in the experiment.
Given an ideal unitary $\hat{\mathcal{U}}_{\textnormal{I}}$ for a two-qubit gate, the average gate fidelity is defined as~\cite{Pedersen2007Fidelity}
\begin{equation}
    F [\hat{\mathcal{U}}_{\textnormal{Q}}] =
    \frac{\Tr [\hat{\mathcal{U}}_{\textnormal{Q}} \hat{\mathcal{U}}^{\dagger}_{\textnormal{Q}}]}{d(d+1)}
    + \frac{ \left|\Tr [\hat{\mathcal{U}}_{\textnormal{Q}} \hat{\mathcal{U}}_{\textnormal{I}}^{\dagger}] \right|^2}{d(d+1)}
\end{equation}
where $\hat{\mathcal{U}}_{\textnormal{Q}}$ is the full unitary truncated to the two-qubit subspace and $d=4$.
Because we ignore the possible single-qubit correction $\hat{Z}\hat{I}$ and $IX$, we compute the maximal fidelity optimized over the possible single-qubit rotation angles
\begin{equation}
    \tilde{F} = \max_{\{\theta_1, \theta_2\}}
    F
    \left[
    \naturalE^{-\im (\theta_1 \hat{I}\hat{X} + \theta_2 \hat{Z}\hat{I})}
    \hat{\mathcal{U}}_{\textnormal{Q}}
    \naturalE^{\im (\theta_1 \hat{I}\hat{X} + \theta_2 \hat{Z}\hat{I})}
    \right]
    .
\end{equation}

\section{Code availability}
The code for calibrating the Cross-Resonance gate can be accessed via this link \url{https://github.com/BoxiLi/qiskit-CR-calibration}.

\begin{acknowledgments}
The authors would like to thank Adrian Lupascu and Francisco Cárdenas López for the insightful discussions.
This work was funded by the Federal Ministry of Education and
Research (BMBF) within the framework programme "Quantum technologies – from
basic research to market" (Project QSolid, Grant No.~13N16149),
by the Deutsche Forschungsgemeinschaft (DFG, German Research Foundation) under Germany's Excellence Strategy – Cluster of Excellence Matter and Light for Quantum Computing (ML4Q) EXC 2004/1 – 390534769,
by HORIZON-CL4-2022-QUANTUM-01-SGA Project under Grant 101113946 OpenSuperQPlus100 and the European Union’s Horizon Programme (HORIZON-CL4-2021-DIGITALEMERGING-02-10) Grant Agreement 101080085 QCFD.
We acknowledge the use of IBM Quantum Platforms for this work. The views expressed are those of the authors, and do not reflect the official policy or position of IBM or the IBM Quantum team. The authors gratefully acknowledge the Gauss Centre for Supercomputing e.V. (www.gauss-centre.eu) for funding this project by providing computing time through the John von Neumann Institute for Computing (NIC) on the GCS Supercomputer JUWELS at Jülich Supercomputing Centre (JSC).
\end{acknowledgments}

\onecolumngrid
\newpage
\begin{center}
  \textbf{\large Supplementary Information for "Experimental error suppression in Cross-Resonance gates via multi-derivative pulse shaping"}\\[.2cm]
\end{center}
\medskip

\setcounter{secnumdepth}{3}
\setcounter{figure}{0}
\setcounter{table}{0}
\setcounter{page}{1}
\setcounter{equation}{0}
\setcounter{section}{0}
\setcounter{table}{0}
\renewcommand{\theequation}{S\arabic{equation}}
\renewcommand{\thefigure}{S\arabic{figure}}
\renewcommand{\bibnumfmt}[1]{[S#1]}
\renewcommand{\thetable}{\arabic{table}}
\renewcommand*{\thesection}{SUPPLEMENTARY NOTE \arabic{section}}
\renewcommand*{\thesubsection}{\Alph{subsection}}

\twocolumngrid

\section{Partial suppression of transition errors}
\label{sec: partial suppression by DRAG}

\begin{figure*}
    \centering
    \includegraphics[width=0.8\linewidth]{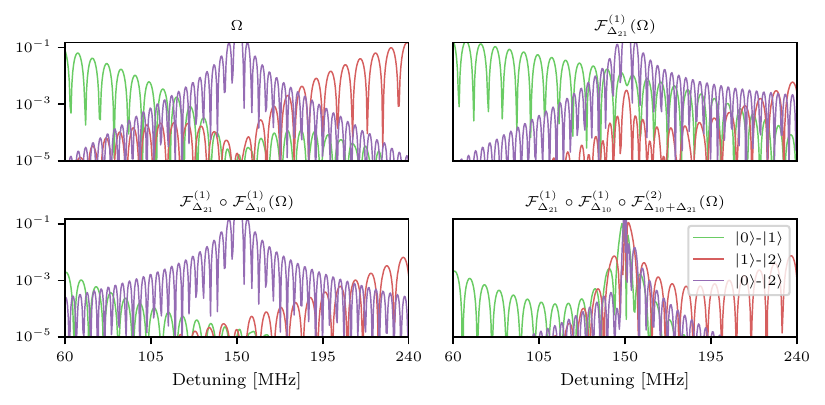}
    \caption{Transition probabilities using different perturbative pulse substitutions.}
    \label{fig:population error step perturbative}
\end{figure*}

\begin{figure*}
    \centering
    \includegraphics[width=0.8\linewidth]{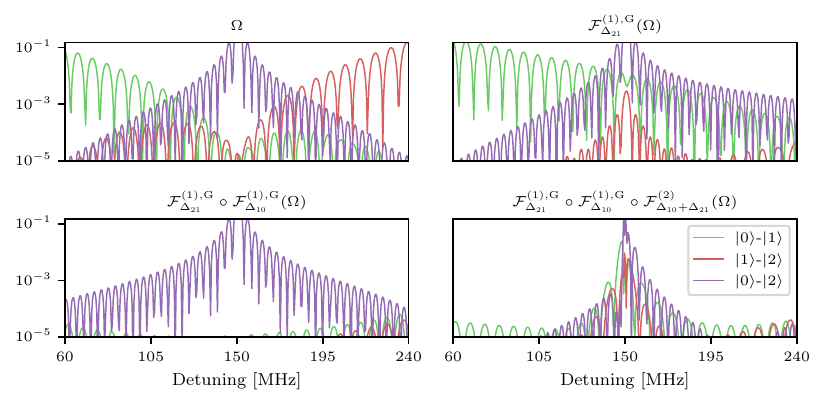}
    \caption{Transition probabilities using the substitutions derived from the exact two-level diagonalization for the single-photon transitions.}
    \label{fig:population error step exact}
\end{figure*}

In \cref{fig:population error step perturbative} and \cref{fig:population error step exact}, we compare the impact of partial suppression of certain errors on the overall performance.
We plot the transition probabilities among the three levels, using pulses designed to suppress none, one, two and all three of the transitions.
The plots illustrate the corresponding suppression of different transitions through the prescribed pulse substitutions.
Additionally, it is evident that a solution targeting only partial suppression inadvertently increases other transitions due to the newly introduced high-frequency components, which underlines the importance of the simultaneous suppression of all transitions.

\section{Robustness of the recursive pulse}
\label{sec:robustness CR drive}
Superconducting qubits often suffer from the drift of the qubit frequency and the drive strength.
In the following, we investigate the performance of the derived analytical pulse shape against those drifts.
For simplicity, we assume that the drift is constant during the CR drive.
We derive the pulse shape using $\Omega_{\max}$ and the control qubit frequency $\Delta_1$ and then perform the two-Transmon simulation with parameter deviations: $\Omega_{\max}+\epsilon_{\Omega}$ and $\Delta_1+\epsilon_{\Delta}$.

\begin{figure}[ht]
    \centering
    \includegraphics[width=\linewidth]{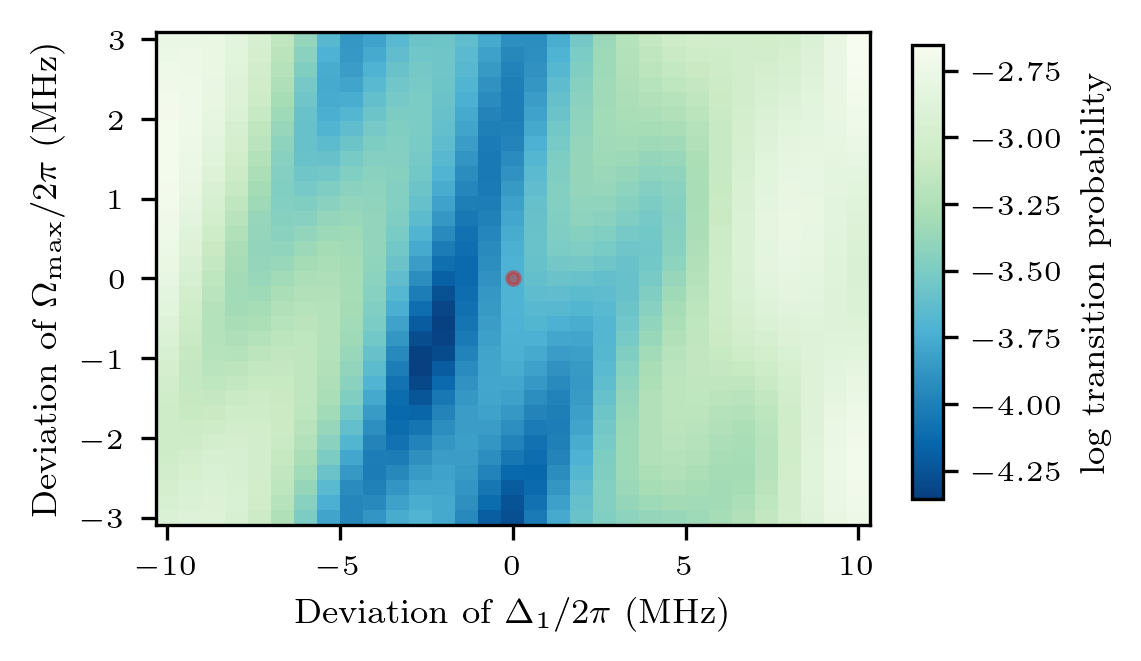}
    \caption{The total error transition probabilities for a precalculated pulse shape under the effect of parameter drift. The red dot marks the data point that uses the initial parameters: $\Omega_{\textnormal{max}}/2\pi=40$~MHz, $\alpha/2\pi=-300$~MHz, $\Delta/2\pi=110$~MHz and $\trise=10$~ns.}
    \label{fig:robustness}
\end{figure}

The total error transition probability is computed from the unitary evolution and depicted on \cref{fig:robustness}.
Although the drift of the drive strength causes some oscillations, it does not significantly increase the error.
A qualitative explanation can be found in the two-level derivation: since the X and Y amplitudes drift simultaneously, the suppression remains the same in the first-order perturbation. 
Only in the next order does it come into the picture through the correction to the energy gap via Stark shift.
The transition error is increased by one order of magnitude if the frequency drifts about 10\% with respect to the qubit-qubit detuning.
Note that in practice drifts are usually much smaller, in the kHz regime.
Not surprisingly, the analytical pulse shape is not located at the region with the absolute lowest error.
Therefore, the performance will benefit from further calibration, both in simulation and experiment. In fact, in the experiment here, we nonetheless do not calibrate these parameters, to demonstrate the remarkable in situ precision of the out-of-the-box pulses.

\section{Amplifying the transition error}
\label{sec:error amplification}
Here we present a simplified derivation of the error amplification technique for the off-resonant error, as discussed in~\cite{Wei2024Characterizing}. In particular, we demonstrate its applicability to multi-photon transitions, such as the $\ket{0}\leftrightarrow\ket{2}$ transition.

For simplicity, we restrict our analysis to a two-level subsystem in which the transition occurs, assuming that other error transitions are not amplified simultaneously. The Hamiltonian is given as
\begin{equation}
\hat{H} = 
\left(
\begin{array}{cc}
 -\frac{\Delta }{2} & \frac{g}{2} \\
 \frac{g}{2} & \frac{\Delta }{2} \\
\end{array}
\right)
.
\end{equation}
The time evolution for a duration $t$, given by $\hat{\mathcal{U}} = \exp(-\im \hat{H} t)$, yields:
\begin{equation}
\hat{\mathcal{U}} = 
    \left(
\begin{array}{cc}
 \cos \left(\frac{t \Delta '}{2}\right)+\frac{\im \Delta  \sin \left(\frac{t \Delta '}{2}\right)}{\Delta '} & -\frac{\im g \sin \left(\frac{t \Delta '}{2}\right)}{\Delta '} \\
 -\frac{\im g \sin \left(\frac{t \Delta '}{2}\right)}{\Delta '} & \cos \left(\frac{t \Delta '}{2}\right)-\frac{\im \Delta  \sin \left(\frac{t \Delta '}{2}\right)}{\Delta '} \\
\end{array}
\right)
\end{equation}
where $\Delta' = \sqrt{(\Delta^2 + \Omega^2)}$.
It clearly shows in the above equation that by prolonging the evolution time $t$, the population error is upper-bounded by $\frac{g^2 \sin^2 \left(t \Delta'/2\right)}{\Delta'^2}$.

According to Ref.~\cite{Wei2024Characterizing}, the transition error can be amplified by introducing a virtual phase gate $\textnormal{RZ}(\phi)$ between the two levels.
The angle $\phi$ is selected to ensure that $\textnormal{RZ}(\phi)\hat{\mathcal{U}}$ induces rotation solely around a fixed axis on the equator, with no rotation around the Z-axis.
Solving the equation under the approximation $\Omega \ll \Delta$ yields $\phi=\Delta't $.

In practice, determining the required angle $\phi$ is not straightforward, requiring a sweep over all possible angles to identify the resonant one.
If the transition is between states $\ket{0}$ and $\ket{1}$, the virtual phase is commonly  implemented by shifting the phase of the drive~\cite{McKay2017Efficient}.
Here, we show that this approach remains valid when the coupling $g$ couples other states such as $\ket{1}$ and $\ket{2}$ or is a multi-photon process like $\ket{0}$ and $\ket{2}$.

To this purpose, we replace $g$ by $\naturalE^{\im n \phi } \frac{\Omega ^n}{\Delta _{\text{eff}}^{n-1}}$ and write the corresponding Hamiltonian
\begin{equation}
\hat{H}_{\Omega}(\theta)
=
\left(
\begin{array}{cc}
 -\frac{\Delta }{2} & \frac{1}{2} \naturalE^{\im n \theta } \frac{\Omega ^n}{\Delta _{\text{eff}}^{n-1}} \\
 \frac{1}{2} \naturalE^{-\im n \theta } \frac{\Omega ^n}{\Delta _{\text{eff}}^{n-1}} & \frac{\Delta }{2} \\
\end{array}
\right)
\end{equation}
where $\Omega$ is the drive amplitude with a phase $\theta$. 
The corresponding unitary evolution is denoted by $\hat{\mathcal{U}}_{\Omega}(\theta)=\naturalE^{-iH_{\Omega}(\theta)t}$.
It is then straight forward to show that $\textnormal{RZ}(\phi) \hat{\mathcal{U}}_{\Omega}(\theta) = \hat{\mathcal{U}}_{\Omega}(\theta-\phi/n)\textnormal{RZ}(\phi)$.
This shows that the virtual phase gate of angle $\phi$ for this two-level subsystem can also be implemented by phase shifting the drive $\Omega$ by $-\phi/n$.
Therefore, both the single photon transitions $\ket{0}\leftrightarrow\ket{1}$, $\ket{1}\leftrightarrow\ket{2}$ and the two-photon transition $\ket{0}\leftrightarrow\ket{2}$ can be amplified by adding virtual phase gate with different $\phi$.
Moreover, sweeping $\phi$ from 0 to $2 \pi$ is expected to reveal two peaks for the two-photon transition.
Although Ref.~\cite{Wei2024Characterizing} suggests preparing the target qubit in the $\ket{+}$ state for a better understanding of the amplification dynamics, we omit it here as we focus solely on the dynamics of the control Transmon.

\section{Data of the used Transmon qubits}
\label{sec:qubits data}

\setlength\heavyrulewidth{0.35ex}
\begin{table*}
\caption{}
\label{tab:ibm_data_default_CR}
\begin{tabular}{@{\quad}lC{1.5cm}C{1.5cm}C{1.5cm}C{1.5cm}C{1.5cm}C{1.5cm}C{1.5cm}@{\,}}
\toprule
\multicolumn{8}{c}{Data of the default CR gate}                                                                                           \\
\midrule
qubit   pairs                         & N (0, 1) & N (2, 1) & N (1, 3) & L (5, 6) & L (3, 1) & L (2, 1) & L (5, 4) \\
control qubit T1 ($\rm{\mu   s}$)     & 109      & 127      & 122      & 158      & 117      & 108      & 158      \\
target qubit T1 ($\rm{\mu   s}$)      & 122      & 122      & 121      & 103      & 62       & 62       & 79       \\
control qubit T2 ($\rm{\mu   s}$)     & 39       & 141      & 82       & 89       & 96       & 93       & 89       \\
target qubit T2 ($\rm{\mu   s}$)      & 82       & 82       & 58       & 80       & 95       & 95       & 30       \\
qubit-qubit detuning $\Delta$   (MHz) & 91       & 104      & 143      & 112      & -113     & 88       & -109     \\
qubit-qubit coupling J (MHz)          & 2.42     & 3.3      & 3.25     & 3.24     & 3.21     & 3.31     & 3.4      \\
default drive amplitude   (a.u.)      & 0.345    & 0.127    & 28.5     & 0.495    & 0.74     & 0.254    & 0.5      \\
estimated amplitude (MHz)             & 70.2     & 19       & 46.6     & 48.8     & 56.7     & 28.7     & 49.3     \\
$\trise$ (ns)                         & 28.4     & 28.4     & 28.4     & 28.4     & 28.4     & 28.4     & 28.4     \\
$ZZ$   (MHz)                       & 0.037    & 0.071    & 0.076    & 0.069    & 0.0.67    & 0.069    & 0.076    \\
$ZX$ CR default (MHz)                   & -1.89    & -0.94    & -1.67    & -1.85    & -1.39    & -1.48    & -1.21    \\
default gate duration (ns)            & 249      & 391      & 270      & 256      & 299      & 291      & 327
\\
\bottomrule
\end{tabular}
\end{table*}

\begin{table*}
\caption{}
\label{tab:ibm_data_custom_CR}
\begin{tabular}{@{\quad}lC{1.5cm}C{1.5cm}C{1.5cm}C{1.5cm}@{\,}}
\toprule
\multicolumn{5}{c}{Data of the self-calibrated CR gate}                       \\ \midrule
qubit pairs                       & N (0, 1) & L (2, 1) & N (2, 1) & L (5, 6) \\
drive amplitude (a.u.)            & 0.345    & 0.5      & 0.4      & 0.6      \\
estimated amplitude (MHz)         & 70.2     & 56.5     & 59.8     & 59.1     \\
$\trise$ (ns)                     & 10       & 10       & 10       & 13       \\
$ZX$   strength (MHz)               & -1.89    & -2.48    & -2.43    & -2.12     \\
direct CNOT duration (ns)         & 146      & 114      & 121      & 135      \\
echoed CNOT duration (ns)         & 227      & 199      & 206      & 220      \\
\bottomrule
\end{tabular}
\end{table*}

In \cref{tab:ibm_data_default_CR} and \cref{tab:ibm_data_custom_CR}, we provide the data of qubits and the parameters used for the CR drive. The coherence time, qubit frequency and the default drive amplitude $J$ were obtained from the IBM Quantum Platform~\cite{IBM} on 17th October 2023. The numbers vary from day to day but in a reasonable range. The effective coupling strength $J$ and the idling $ZZ$ strength were also obtained from the default calibration data.

For the default CR gate, the effective Hamiltonian terms and the default gate time are measured during our calibration procedure.
For the self-calibrated CR gate, the drive amplitude is chosen empirically. Due to our limited access, we are unable to sweep through different drive amplitudes and optimize for the optimal gate time.
Instead, the drive strength and $\trise$ are chosen empirically based on our knowledge of the qubit-qubit detuning and the saturation of the effective $ZX$ coupling.
Further improvements are expected through a more comprehensive and in-depth calibration.

\section{Additional data on the transition error suppression}
\label{sec: comparison drag schemes}

\begin{figure}[th]
    \centering
    \includegraphics[width=\linewidth]{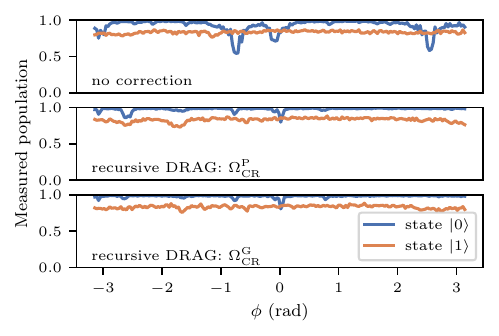}
    \caption{The transition error under different pulse schemes for a pair of qubits with $
    \Delta_{10}=143$~MHz. The data is obtained from {\it ibm\_nairobi} Q1 and Q3 with the drive amplitude is 0.3 ($\approx 51$~MHz), $\trise=15$~ns and $N=60$.}
    \label{fig:drag calibration nairobi13}
    \centering
    \includegraphics[width=\linewidth]{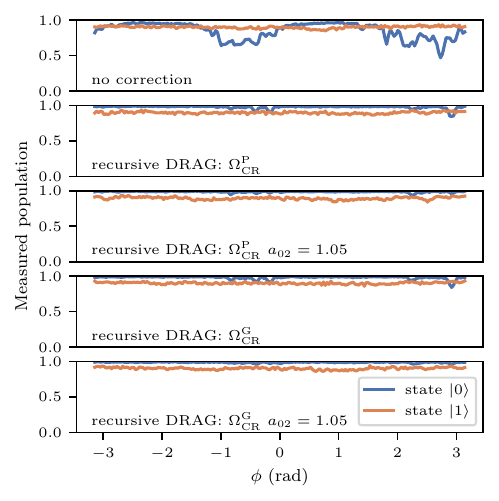}
    \caption{The transition error under different pulse schemes for a pair of qubits with $
    \Delta_{10}=189$~MHz. The data is obtained from {\it ibm\_lagos} Q5 and Q3 with the drive amplitude is 0.5 ($\approx 49$~MHz), $\trise=15$~ns and $N=30$.}
    \label{fig:drag calibration lagos53}
\end{figure}

\begin{figure}[th]
    \centering
    \includegraphics[width=\linewidth]{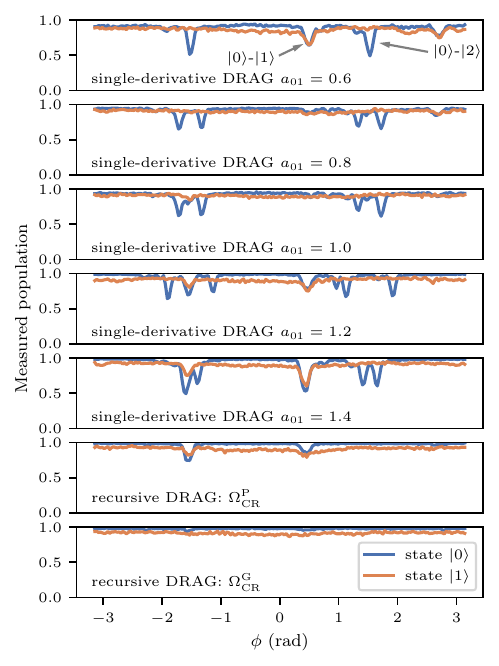}
    \caption{Comparison between different \textsc{drag} schemes. Plotted are the amplified transition errors of single-derivative \textsc{drag} pulses with varying \textsc{drag} coefficients, alongside the two recursive \textsc{drag} pulses proposed in this paper.
    Data is obtained from {\it ibm\_lagos} Q5 and Q6, with $\Delta=112$~MHz, $\tr=10$~ns and a drive amplitude about 40 MHz.
    }
    \label{fig:comparison drag experiment}
\end{figure}

In the following, we show additional data on the validation of the recursive \textsc{drag} pulse for suppressing the transition errors on the control qubit.
We compare it to the single-derivative \textsc{drag} pulse used in previous experiments and discuss the calibration of multiple DRAG parameters.

In addition to the qubit pair with the qubit-qubit detuning 104~MHz shown in \cref{fig:transition error experiment}, we show two pairs of qubits with detuning 143~MHz and $189$~MHz in \cref{fig:drag calibration lagos53,fig:drag calibration nairobi13}.
In both cases, the recursive \textsc{drag} demonstrates excellent performance without any further calibration.
Because the single-photon transition error is not very large, there is little difference between the perturbative solution and the Gives rotation in those two cases.

In \cref{fig:population error}b, we compare the performance of the single-derivative \textsc{drag} and the proposed recursive \textsc{drag} methods through simulation.
In \cref{fig:comparison drag experiment}, we show an example where, despite the calibration of the \textsc{drag} coefficient, the single-derivative \textsc{drag} fails to sufficiently suppress all errors, whereas the proposed methods exhibit excellent performance.
We plot the amplified error for different \textsc{drag} coefficients with the single-derivative \textsc{drag} scheme with a free parameter $a_{01}$. Evidently, while the $\ket{0}\leftrightarrow\ket{1}$ transition can be sufficiently suppressed with a properly chosen \textsc{drag} coefficient, other errors, such as the $\ket{0}\leftrightarrow\ket{2}$ transitions, remain largely unaffected.
In contrast, both recursive methods show substantial improvement, with the pulses derived by Givens rotation achieving a perfect suppression up to the resolution of our amplification circuits, consistent with the performance observed in the qubit pairs illustrated in \cref{fig:population error}e.

Although the recursive \textsc{drag} pulse needs little calibration for this problem, in some scenarios, especially only perturbative \textsc{drag} is used, calibration of the \textsc{drag} parameter may still prove useful.
This can be achieved by adding a free parameter before each correction term in the substitution formula.
In \cref{fig:drag calibration lagos53}, we replace the substitution for the $\ket{0}\leftrightarrow\ket{2}$ transition in \cref{eq:explicit perturbative recursive DRAG pulse} to 
\begin{equation}
    \Omega_2 = 
    \sqrt{
        \Omega_3^2 - \im a_{02}\frac{2\Omega_3\dot{\Omega}_3}{\Delta_{20}}
        }
        .
\end{equation}
By varying the free parameter $a_{02}$, the $\ket{0}\leftrightarrow\ket{2}$ transition can be fine-tuned.
Thanks to its recursive structure, the suppression of other transitions remains unaffected.

\begin{figure}[th]
    \centering
    \includegraphics[width=0.8\linewidth]{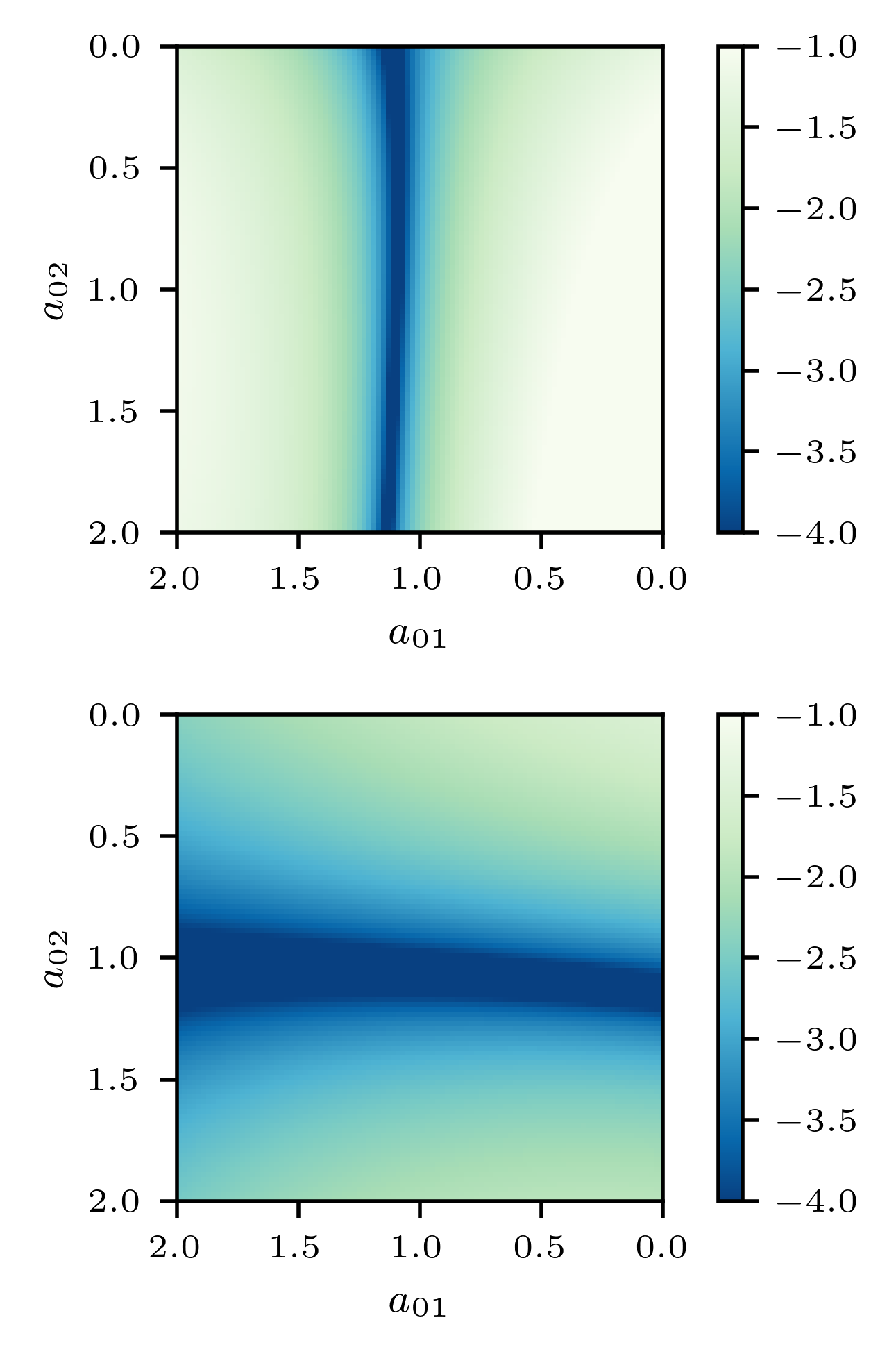}
    \caption{Transition error probability between $\ket{0}\leftrightarrow\ket{1}$ (top) and $\ket{0}\leftrightarrow\ket{2}$ (bottom) as a function of two different \textsc{drag} coefficients. The \textsc{drag} coefficient affects the targeted error but has only little effect on the other one. The parameters used in the simulation are $\alpha=-300$~MHz, $\Delta_{10}=110$~MHz, $\Omega_{\max}=50$~MHz and $\trise=12$~ns.}
    \label{fig:cr drag coeff sweep 2d}
\end{figure}

This independence between different parameters is illustrated more clearly in the simulation result in \cref{fig:cr drag coeff sweep 2d}.
Here we also modify the perturbative \textsc{drag} substitution for the $\ket{0}\leftrightarrow\ket{1}$ transition to 
\begin{equation}
    \Omega_{\textnormal{CR}}^{\rm{P}} = \Omega_1 - \im a_{01} \frac{\dot{\Omega}_1}{\Delta_{10}}
    .
\end{equation}
We simulate the dynamics of the three-level Hamiltonian introduced in the main text.
By sweeping the \textsc{drag} parameters $a_{01}$ and $a_02$, we obtain the transition error probabilities shown in \cref{fig:cr drag coeff sweep 2d}.
The calibration of one of the \textsc{drag} parameters has little effect on the other.
Thus, the two parameters can be calibrated independently with a few iterations without a full two-dimensional sweep.

\section{Calibration of the CNOT gate}
In the following, we detail the calibration of the CNOT gate using CR interaction.
Our calibration routine is based on the default calibration data of the Transmon frequency, anharmonicity and single-qubit X gate on the IBM Quantum Platform.
\label{sec:cr calibration}
\subsection{Hamiltonian tomography}

\begin{figure}[t]
\centering
\includegraphics[width=\linewidth]{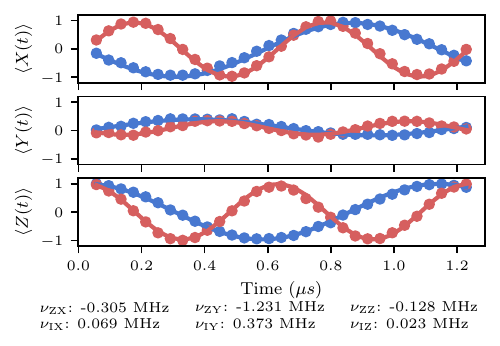}
\includegraphics[width=\linewidth]{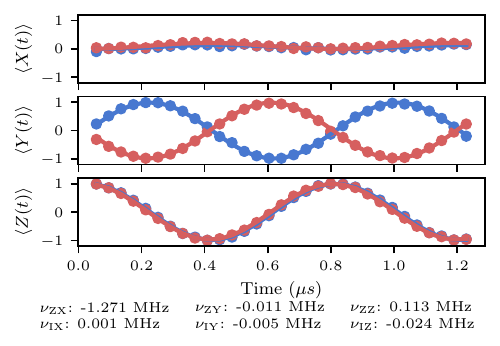}
\caption{
Experimental Hamiltonian tomography data of the CR pulse, before (top panel) and after (bottom panel) the calibration. The blue (red) colour corresponds to the dynamics of the target qubits under the CR drive when the control qubit is in $\ket{0}$ ($\ket{1}$).
}
\label{fig:hamiltonian tomography}
\end{figure}

We present here the Hamiltonian tomography used in calibrating the CR pulse, based on Ref.~\cite{Sheldon2016Procedure,Danin2023Procedure} and the qiskit online tutorial.
We start by the CR Hamiltonian in the effective frame~\cref{eq:effective hamiltonian cr}.
Calibrating a CR gate involves measuring the coefficients $\nu$ and eliminating undesired terms, achieved by selecting an appropriate phase for the CR drive and applying a compensatory drive on the target qubit.
An example of the tomography data is shown in \cref{fig:hamiltonian tomography}.

To begin with, we observe that in \cref{eq:effective hamiltonian cr}, the CR dynamics involve the rotation of the target qubit, depending on the state of the control qubit.
Therefore, characterization can be achieved by conducting single-qubit Hamiltonian tomography on the target qubit while preparing the control qubit in states $\ket{0}$ and $\ket{1}$.

In the following, we derive the equations that are used to fit the measured data in \cref{fig:hamiltonian tomography} and obtain the coefficients $\nu$. In the Heisenberg picture, the time evolution of an observable $\hat{O}$ is given by
\begin{equation}
    \frac{\dd \hat{O}}{\dd t} = \im [\hat{H}, \hat{O}]
    \label{eq:Heisenberg picture}
    .
\end{equation}
Measuring the target qubits on different bases yields the expectation values $\expect{\hat{I}\hat{X}}$, $\expect{\hat{I}\hat{Y}}$ and $\expect{\hat{I}\hat{Z}}$.
Plugging these into \cref{eq:Heisenberg picture} results in the following expressions:
\begin{align}
    &
    \frac{\dd}{\dd t} \hat{I}\hat{X}
    =
    \left(
    \nu_{ZY} \hat{Z}\hat{Z}
    -\nu_{ZZ} \hat{Z}\hat{Y}
    +\nu_{IY}  \hat{I}\hat{Z}
    -\nu_{IZ}  \hat{I}\hat{Y}
    \right)
    \nonumber\\
    &
    \frac{\dd}{\dd t} \hat{I}\hat{Y}
    =
    \left(
    -\nu_{ZX} \hat{Z}\hat{Z}
    +\nu_{ZZ} \hat{Z}\hat{X}
    -\nu_{IX}  \hat{I}\hat{Z}
    +\nu_{IZ}  \hat{I}\hat{X}
    \right)
    \nonumber\\
    &
    \frac{\dd}{\dd t} \hat{I}\hat{Z}
    =
    \left(
    +\nu_{ZX} \hat{Z}\hat{Y}
    -\nu_{ZY} \hat{Z}\hat{X}
    +\nu_{IX} \hat{I}\hat{Y}
    -\nu_{IY} \hat{I}\hat{X}
    \right)
    .
\end{align}
Assuming that the control qubit is prepared in a computational basis and remains unchanged during the evolution, these equations can be further simplified with the expectation values on the target qubit:
\begin{align}
    &\frac{\dd}{\dd t} \expect{\hat{X}}
    =
    \left(b \nu_{ZY} + \nu_{IY}\right) \expect{\hat{Z}} - \left(b \nu_{ZZ} + \nu_{IZ}\right) \expect{\hat{Y}}
    \nonumber\\
    &\frac{\dd}{\dd t} \expect{\hat{Y}}
    =
    -\left(b \nu_{ZX} + \nu_{IX}\right) \expect{\hat{Z}} + \left(b \nu_{ZZ} + \nu_{IZ}\right) \expect{\hat{X}}
    \nonumber\\
    &\frac{\dd}{\dd t} \expect{\hat{Z}}
    =
    \left(b \nu_{ZX} + \nu_{IX}\right) \expect{\hat{Y}} - \left(b \nu_{ZY} + \nu_{IY}\right) \expect{\hat{X}}
    ,
\end{align}
where $b=1$ ($b=-1$) if the control qubit is in state $\ket{0}$ ($\ket{1}$).
To further simplify the notation, we define $\omega_X^{(b)}=b \nu_{ZX} + \nu_{IX}$ and the same for $Y$ and $Z$ axis.

The differential equation can be solved by exponentiating the generator
\begin{equation}
G^{(b)}=\left(
\begin{array}{ccc}
 0 & -\omega _Z & \omega _Y \\
 \omega _Z & 0 & -\omega _X \\
 -\omega _Y & \omega _X & 0 \\
\end{array}
\right)
\end{equation}
where we omit the upper script $(b)$ of $\omega$ for simplicity.
Providing that the target qubit is always initialized in the ground state, the solution is as follows:
\begin{align}
    &\expect{\hat{X}(t)}
    =
    \frac{1}{\omega ^2}
    \left(
    -\omega _X \omega _Z \cos (t \omega )+\omega  \omega _Y \sin (t \omega )+\omega _X \omega _Z
    \right)
    \nonumber \\
    &\expect{\hat{Y}(t)}
    =
    \frac{1}{\omega ^2}
    \left(
    -\omega  \omega _X \sin (t \omega )-\omega _Y \omega _Z \cos (t \omega )+\omega _Y \omega _Z
    \right)
    \nonumber \\
    &\expect{\hat{Z}(t)}
    =
    \frac{1}{\omega ^2}
    \left(
    (\omega _X^2 +\omega _Y^2 )\cos (t \omega )+\omega _Z^2
    \right)
    ,
\end{align}
where $\omega = \sqrt{\omega _X^2+\omega _Y^2+\omega _Z^2}$.
These equations are then used to fit the measured data.

Although the above derivation is based on a constant drive pulse, it applies also to our time-dependent pulses used in this work.
During a tomography experiment, the time-dependent pulse ramping period is fixed while the holding time is adjusted from zero to a maximal duration.
The tuning-up (-off) period of the CR drive, brings the system into (out of) the effective frame, while the tomography assesses the dynamics of the holding period.
Additionally, the variation in the constant phase of the drive pulse merely affects the rotation axis of the target qubit without changing the underlying dynamics.

In practical applications, fitting trigonometric functions with undetermined oscillation frequencies can be challenging, depending heavily on the initial values. Therefore, an iterative fitting procedure is employed. The dynamics of $\expect{\hat{Z}(t)}$ are first fitted to obtain a good estimation of $\omega$. Then, the other two equations are included one by one, forming an iterative fitting process.
In addition, it is helpful to not force the renormalization $\omega _X^2+\omega _Y^2+\omega _Z^2 = \omega ^2$ at the beginning.
Instead, it is used to fine-tune the result in later stages, leveraging the previous values as an initial guess.

\subsection{Calibration of the echoed CNOT gate}
The calibration process for the echoed CNOT gate involves three main steps:
\begin{enumerate}
    \item Adjusting the phase of the CR drive and calibrating the target compensation drive. This step ensures that the $ZY$, $IX$ and $IY$ terms are removed from the effective Hamiltonian in \cref{eq:effective hamiltonian cr}.
    \item Calibrating the $IY$-\textsc{drag} amplitude and determining the pulse detuning. In this step, three different $IY$-\textsc{drag} amplitudes are sampled. The zero points of the $ZZ$ coupling strength are determined through a linear fit (see \cref{fig:ZZ IZ correction}). Simultaneously, the measured $IZ$ coefficient provides information about the detuning of the two drives.
    \item Computing the pulse duration from the tomography data. In particular, for an echoed CNOT gate, each CR pulse should be configured to last one-eighth of the period (see \cref{fig:hamiltonian tomography}), since the target qubit rotates towards the opposite direction depending on the state of the control. This ensures the generation of a precise 90-degree $ZX$ rotation.
\end{enumerate}

While the second step requires only a linear fit, calibration of the CR and target drive in step one cannot be completed in one round in many cases because of the nonlinearity.
Therefore, we iterate a few steps until the unwanted terms are suppressed below a certain threshold.
In the following, we derive the update function of one calibration step.

We first define the following notation
\begin{align}
\Omega_{\rm{CR}}=|\Omega_{\rm{CR}}|\naturalE^{\im \theta_1} &\coloneqq \Omega_{\rm{CRX}} + \im \Omega_{\mathrm{CRY}} \\
\Omega_{\rm{CRX}}=|\Omega_{\rm{T}}|\naturalE^{\im \theta_2} &\coloneqq \Omega_{IX} + \im \Omega_{IY}
.
\end{align}
The notation introduced provides a clear separation of amplitude and phase in the pulse design.
It's important to note that the time dependence (pulse shape) is not included in this definition and is not changed during the calibration.
Here, $\Omega$ represents only the maximal amplitude and a constant phase of the pulse.

The iterative calibration process begins with a predefined $|\Omega_{\rm{CR}}|$, with $\theta_1$, $|\Omega_{\rm{T}}|$ and $\theta_2$ all set to zero, to be updated iteratively.
At each iteration $k$, we perform two tomography experiments.
The first tomography is performed with the calibrated parameters from the previous step and measures different coupling coefficients $\nu$ of the Hamiltonian
\begin{equation}
    \hat{H}(\Omega_{\rm{CR}}, \Omega_{\rm{T}}) = \nu_{ZX}\hat{Z}\hat{X} + \nu_{ZY}\hat{Z}\hat{Y} + \nu_{IX}\hat{I}\hat{X} + \nu_{IY}\hat{I}\hat{Y}
    .
\end{equation}
Here we omit the $ZZ$ and $IZ$ terms as they are not the target in this calibration.
If only the $ZX$ term is significant and all other three terms small enough, the calibration terminates.

Given the first tomography, the phase of the CR drive for the $k+1$ iteration can be easily adjusted by 
\begin{equation}
\theta_1^{(k+1)} = \theta_1 - \arctan{\frac{\nu_{ZY}}{\nu_{ZX}}}
\end{equation}
where on the right-hand side we omit the upper index for step $k$.

To calibrate the compensation target drive, a second tomography experiment is performed with a different $\Omega_{\textnormal{T}}'$ and results in the following Hamiltonian
\begin{equation}
    \hat{H}(\Omega_{\rm{CR}}, \Omega_{\rm{T}}') = \nu_{ZX}\hat{Z}\hat{X} + \nu_{ZY}\hat{Z}\hat{Y} + \nu_{IX}'\hat{I}\hat{X} + \nu_{IY}'\hat{I}\hat{Y}
\end{equation}
with $\Omega_{\rm{T}}'=|\Omega_{\rm{T}} + \Delta\Omega|\naturalE^{\im \theta_2}$, introducing a change $\Delta\Omega$ in the drive amplitude.
Note that the coefficients of the coupling terms do not change because we only changed the target drive amplitude. 
This step is crucial for precisely calibrating the compensation target drive because in the qiskit user interface, the amplitude $\Omega_{\rm{T}}$ is defined in a renormalized arbitrary unit from zero to one.

With the above two tomography experiments, the new amplitude $\Omega^{(k+1)}_{\rm{T}}$ and phase $\theta^{(k+1)}_2$ of the target drive can be computed as follows.
First, the difference between the two measured Hamiltonian yields
\begin{equation}
    \nu_{\textnormal{T}} - \nu_{\textnormal{T}}' = \left(\Omega_{\textnormal{T}} - \Omega_{\textnormal{T}}'\right)\naturalE^{\im \theta_2}C_{\rm{T}} \naturalE^{-\im \phi_T}
\end{equation}
where $\nu_{\textnormal{T}} = \nu_{IX} + \im \nu_{IY}$ and $\nu_{\textnormal{T}}' = \nu_{IX}' + \im \nu_{IY}'$.
This follows from the assumption that locally the drive amplitude and the coefficients of the effective Hamiltonian show a linear relation characterized by $C_{\rm{T}} \naturalE^{-\im \phi_T}$.
Similarly, for the desired effective Hamiltonian terms with the coefficients denoted by $\nu_{\rm{T, \rm{ideal}}}$, we have
\begin{equation}
    \nu_{\rm{T, \rm{ideal}}} - \nu_{\textnormal{T}} = 
    \left(
    \Omega^{(k+1)}_{\mathrm{T}}e^{\im \theta^{(k+1)}_2} - \Omega_{\textnormal{T}}\naturalE^{\im \theta_2}
    \right)
    C_{\rm{T}} \naturalE^{-\im \phi_2}
    .
\end{equation}
The solution is given by:
\begin{equation}
    \Omega^{(k+1)}_{\mathrm{T}}e^{\im \theta^{(k+1)}_2} =
    \Omega_{\textnormal{T}}\naturalE^{\im \theta_2}
    +
    \frac{\nu_{\rm{ideal}} - \nu}{\nu - \nu'}(\Omega_{\textnormal{T}} - \Omega_{\textnormal{T}}')\naturalE^{\im \theta_2}
    .
    \label{eq:cr_calibration_solution}
\end{equation}
This equation provides the updated parameters for the next iteration in the calibration process.
If $\theta_1^{(k+1)}$ is updated, the update $\theta_1^{(k+1)}-\theta_1^{(k)}$ must also be added to the target drive $\theta_2^{(k+1)}$.

\subsection{Calibration of the direct CNOT gate}

\begin{figure}[t]
    \centering
    \includegraphics[width=\linewidth]{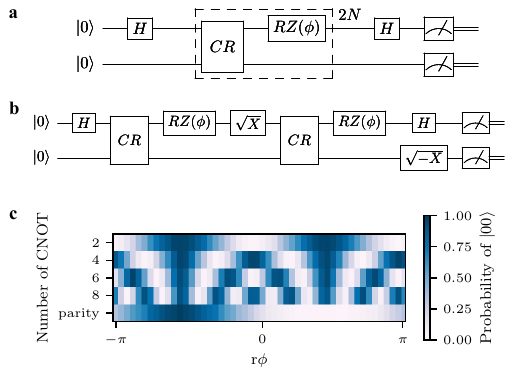}
    \caption{Calibration of the Z phase on the control qubit. (a) Angle calibration circuit for the Z phase. (b) Verification circuit for the Z phase. (c) Example of the calibration data. The first 4 rows correspond to the circuit a while the last row is the result of the circuit b.}
    \label{fig:rz0_calibration}
\end{figure}

The calibration process for the direct CNOT gate is based on the echoed CNOT calibration and involves two additional steps.

First, we adjust the target drive such that $\nu_{IX} = \nu_{ZX}$.
Essentially, we aim for the target qubit to rotate exclusively when the control is in the state $\ket{1}$.
This can be easily implemented with the previously introduced iterative calibration process by setting $\nu_{\rm{ideal}}=\nu_{ZX}$ in \cref{eq:cr_calibration_solution}.
It is noteworthy that, in this case, the tomography experiment with the control qubit in state $\ket{0}$ yields minimal information and can be omitted.

Following the target drive calibration, the next step involves calibrating the phase shift on the control qubit.
This phase shift is caused by the Stark shift induced by the CR drive.
Unlike the echoed gate, where the phase accumulated is automatically removed by the echoing configuration, the direct gate requires explicit calibration of this phase shift.
To accomplish this, we employ the circuits depicted in \cref{fig:rz0_calibration}a and b.
The first circuit in \cref{fig:rz0_calibration}a applies $2N$ uncalibrated CR gate, each combined with a $RZ(\phi)$ rotation on the control.
This gate sequence is sandwiched by Hadamard gates to measure the accumulated phase.
The qubits return to the initial state only if the CR gate combined with the rotation gives a CNOT or a CNOT with a 180-degree rotation.
To select the correct result, we use the verification circuit depicted in \cref{fig:rz0_calibration}b, which returns to the original state only for the correct CNOT gate.
An example of the calibration data is shown in \cref{fig:rz0_calibration}c.

\end{document}